\documentclass[aps,prd,twocolumn,nofootinbib,longbibliography,10pt]{revtex4}
\usepackage{amsmath,amssymb,amsfonts,mathtools,units}
\usepackage{graphicx}
\usepackage[colorlinks=true,urlcolor=blue,citecolor=red,linkcolor=blue]{hyperref}
\usepackage{tensor}
\usepackage{natbib}
\usepackage{flushend,BOONDOX-cal,BOONDOX-frak}
\DeclarePairedDelimiter{\abs}{\lvert}{\rvert}

\makeatletter
\newsavebox{\@brx}
\newcommand{\llangle}[1][]{\savebox{\@brx}{\(\m@th{#1\langle}\)}%
  \mathopen{\copy\@brx\kern-0.5\wd\@brx\usebox{\@brx}}}
\newcommand{\rrangle}[1][]{\savebox{\@brx}{\(\m@th{#1\rangle}\)}%
  \mathclose{\copy\@brx\kern-0.5\wd\@brx\usebox{\@brx}}}
\makeatother

\begin{document}
\title{\textbf{Renormalization group improved cosmology in the presence of a stiff matter era}}
\author{Gopinath Guin}
\email{gopinathguinphysics@gmail.com}
\affiliation{Department of Astrophysics and High Energy Physics, S. N. Bose National Centre for Basic Sciences, JD Block, Sector-III, Salt Lake City, Kolkata-700 106, India}
\author{Soham Sen}
\email{sensohomhary@gmail.com}
\affiliation{Department of Astrophysics and High Energy Physics, S. N. Bose National Centre for Basic Sciences, JD Block, Sector-III, Salt Lake City, Kolkata-700 106, India}
\author{Sunandan Gangopadhyay}
\email{sunandan.gangopadhyay@gmail.com}
\affiliation{Department of Astrophysics and High Energy Physics, S. N. Bose National Centre for Basic Sciences, JD Block, Sector-III, Salt Lake City, Kolkata-700 106, India}
%%%%%%%%%%%%%%%%%%%%%%%%%%%%%%%%%%%%%%%%%%%%%%%%%%%%%%
\begin{abstract}
\noindent In \href{https://link.aps.org/doi/10.1103/PhysRevD.92.103004}{Phys. Rev. D 92 (2015) 103004}, simple analytical solutions of the Friedman equations were obtained for a universe having stiff matter component in the early universe together with a dark matter, and a dark energy component. In this analysis, the universe is considered to be made of a dark fluid which behaves as a stiff matter in the early phase of the universe (when the internal energy dominates). It is also more logical to consider quantum gravitational effects in the early phase of the cosmological evolution. In this analysis, following \href{https://link.aps.org/doi/10.1103/PhysRevD.65.043508}{Phys. Rev. D 65 (2002) 043508}, we consider renormalization group improved modified Friedmann equations where the Newton's gravitational constant ($G$) and the cosmological constant ($\Lambda$) flows with the momentum scale $k$ of the universe. In \href{https://link.aps.org/doi/10.1103/PhysRevD.65.043508}{Phys. Rev. D 65 (2002) 043508}, $k$ is identified with either $\nicefrac{1}{t}$ or with the inverse of the scale factor of the universe $\left(\nicefrac{1}{a(t)}\right)$. In this work, we obtain the momentum scale in terms of the scale factor of the universe when a stiff-matter era is present in the early time regime ($t$ being smaller than the Planck time). It is observed that for a universe undergoing a stiff matter era, radiation era, and matter era, inflation is absent in the early time regime of the universe when the flow of the Newton's gravitational constant and cosmological constant is under consideration. Using the identification of the momentum scale with the scale factor of the universe, we then explore the era $t>t_{\text{Pl}}$ which indicates a primarily matter dominated era with accelerated expansion due to the presence of dark energy. Finally, considering the total equation of state as a combination of linear equation of state along with a polytropic equation of state, we observe that after the Planck-time the universe can undergo an inflationary phase and we find out that the inflation is enhanced by quantum gravitational effects coming due to consideration of renormalization group improvement in the Newton's gravitational constant as well as the cosmological constant. Finally, we propose a new energy density that shows the transition between a primordial stiff matter era followed by an inflation era just after the Planck time, and then the transition from inflation to radiation dominated reheating phase around a second transition point in this renormalization group improved cosmological model.
\end{abstract}
%\arxivnumber{2411.03693}
\maketitle
\section{Introduction}
\noindent The initial notion of the universe was that it is homogeneous, isotropic and static at the same time. Einstein in \cite{AEinstein1,AEinsteinReview}, used his general relativistic field equations and applied it to cosmology. He took into account the existence of homogeneity and isotropy in the universe. He has considered a static universe which was the standard notion upto then where the universe had no beginning and it exists forever. In order to obtain a static solution he needed to consider a cosmological constant in the field equations. In the same year de Sitter also proposed a static cosmological solution with the astrophysical bodies and system considered to be test particles, in capable of affecting the geometry of the spacetime \cite{deSitter1}. In his paper, he predicted that the light emitted from the sources will be redshifted because of the repulsive nature generated due to the cosmological constant and with increasing distance time will slow down which is known as the de Sitter's effect. The static model of Einstein could not explain redshift whereas de-Sitter's model consisted of no matter leading to certain inconsistencies. First Friedmann \cite{Friedmann1,Friedmann2} and then Lanczos \cite{Lanczos}, Weyl \cite{Weyl}, and Lema\^{i}tre \cite{Lemaitre,Lemaitre2} proposed non-stationary solutions of the universe. This particular model  of the universe explained red-shift without resorting to any adhoc assumptions. Robertson also gave a reinterpretation of the de Sitter solution by taking dynamics into consideration \cite{Robertson1,Robertson2}. In \cite{Eddington}, Eddington argued and proved that the static model of the universe proposed by Einstein is unstable which was also supported by the observations made by Edwin Hubble \cite{Hubble}. As a result it was evident that the universe is homogeneous, isotropic but non-static in nature. Finally, in \cite{Lemaitre3,Lemaitre4}, Lema\^{i}tre proposed that the universe must have started from a singular point (a singularity) as it is expanding. This paved the way for the Big Bang model of the universe. One problem with the Big Bang model is that at $t=0$ the energy density goes to infinity. This is also known as the singularity problem of the universe. Such problems only occur because at such high energy scale, quantum gravity effects should come into effect rather than general classical field equations. The Big Bang theory has several problems like the flatness problem, horizon problem, and monopole problem which can be solved using a inflation in the evolution of the cosmos \cite{RGCosmo0,RGCosmo1,RGCosmo2}\footnote{One can also look into non-singular bouncing cosmological models for a physical analysis of spacetime defects \cite{Battista}.}. One of the ways to deal with the horizon problem and the flatness problem is by introducing renormalized group improved cosmology. Renormalization group improvement in black hole spacetime was first done in \cite{BonanoReuter99,BonanoReuter}. In \cite{BonanoReuter2}, renormalization group technique was implemented to determine the leading order quantum gravity corrections in the well known Friedmann-Lema\^{i}tre-Robertson-Walker (FLRW) cosmology. The basic step is to consider an effective average action $\Gamma_k[g_{\mu\nu}]$ where $k$ is a momentum scale with $\Gamma_k$ perfectly taking care of all tree level gravitational phenomenon as well as all loop corrections (for momentum of the order $k$). If $\Gamma_k$ is treated as a function of the momentum scale $k$, then $\Gamma_k$ implies a renormalization group trajectory which is placed in the space of all action functionals. One can determine such a trajectory by solving the ``\textit{flow equations}". Non-perturbative approximate solutions for the renormalization group equations  can be obtained via the method of ``\textit{truncation}" where from the infinite dimensional space of all action functionals, one projects the renormalization group flow onto some finite dimensional relevant subspace. This \textit{truncation} method converts the functional renormalization group equations to a system of ordinary differential equations where the ordinary differential equations involves a finite set of generalized coupling constants. For a renormalization group improved FLRW cosmological model, the generalized coupling constants are the Newton's gravitational constant $G(k)$ and cosmological constant $\Lambda(k)$ where the coupling constants depend on the momentum scale $k$. In \cite{Reuter}, the differential equations governing the dynamics of $G(k)$ and $\Lambda(k)$ depending on the momentum scale $k$ were derived and the solutions were further discussed in \cite{Souma,BonanoReuter}. If one takes a very high value of $k$ ($k\rightarrow\infty$), then the dimensionless Newton's gravitational constant $g(k)=k^2 G(k)$ approaches a $UV$ attractive non-Gaussian fixed point $g_*^{UV}$ which also implies that gravity is asymptotically free ($G(k)\rightarrow 0$ for $k\rightarrow\infty$). The Reuter fixed point in $2+\epsilon$-dimensions is similar to the current fixed point for 3+1-dimensional quantum gravity theory \cite{Weinberg0}. The renormalization group improvement is implemented by replacing $G$ by $G(k)$ and $\Lambda$ by $\Lambda(k)$ in the Einstein's field equations. In \cite{BonanoReuter2} it was argued that the renormalization group improved solutions are ``\textit{natural}" implying the absence of the need of any fine tuning and the solutions are free of particle horizons for a broad class of equations of state. Therefore, these solutions get rid of the flatness problem and horizon problem without the need of an inflation era. There also has been several literature based on renormalized group improved cosmology in recent times \cite{RGCosmo0,RGCosmo1,RGCosmo2}. 

\noindent Recently in \cite{Donoghue}, it was argued that the asymptotic safety approach discussed above has some fundamental flaws. It was argued in \cite{Donoghue} that the asymptotic safety approch runs into some conflicts with exact analytical results in low energies. It has been argued that just by applying renormalization group improvement to Newton's coupling the one-loop correction to gravity mediated scattering amplitude cannot be obtained and one needs to improve the ``standard practice" of renormalization group improvement by implementing a Lorentzian signature instead of using the Euclidean signature. As has been discussed in \cite{ReplytoDonoghue}, the implementation of a metric with Euclidean signature has the benefit of easily defining the direction of the Renormalization group flow. For a quantum field theory in flat spacetime one calculate the propagators in ther Euclidean signature and then restore the Lorentzian model by implementing a Wick rotation of the Euclidean time. In curved spacetime however, the analytical continuation from the Euclidean to Lorentzian signature is not straighforward and rather has some very severe obstacles. One of them being the complex structure of the graviton propagator which will restrict one to use the analytical continuation from Euclidean to Lorentzian signature. The momentum regulation in this process may break the global spacetime symmetries or result in generating new poles \cite{AnalyticalContinuationIssue}. Hence the extension of analytical continuation in asymptotic safety gravity has not yet been implemented. Hence, we stick with the standard practice of this functional renormalization-group approach and will use the standard techniques thoroughly used in several literatures.

\noindent It is important to note all of the analyses above has considered the existence of a radiation era, a dark matter era. The current accelerated expansion of the universe is governed by a dark energy dominated era, also known as the late inflation. The cosmological model first proposed by Zel'dovich \cite{Zeldovich}, brought in the possibility of a primordial stiff matter era. In this model, the primordial universe is assumed to be constructed of a very cold gas of baryons governed by an equation of state $P=\varepsilon$. He showed that for such a stiff matter era the speed of sound is almost equal to the speed of light \cite{Zeldovich2}. As the energy density corresponding to the stiff matter era $\varepsilon_{\text{s}}$ is proportional to $a(t)^{-6}$, it will dominate before the radiation, dark matter, and dark energy era. After the success of the hot big bang model the model proposed by Zel'dovich was abandoned although, as argued in \cite{NatureStiff}, a primordial stiff matter era is of great physical interest. A stiff-matter era implicates the initial state of the universe was in a state of inactivity rather than in a state of chaos \cite{NatureStiff}. For relativistic scalar field theory, if the kinetic term dominates the potential term of the relativistic scalar field, the scalar field starts to behave as a stiff fluid. In \cite{BECSF}, a fully relativistic treatment of Bose-Einstein condensate dark matter/ stiff fluid was conducted which explicitly showed the existence of a primordial stiff matter era, followed by a radiation era and a matter era where the radiation era existed because of a self-interacting stiff fluid. Using a hydrodynamical representation of the Klein-Gordon-Einstein equation the same model proposed in \cite{BECSF} was analyzed in \cite{ChavanisSuarez}. In the same year  in \cite{Chavanis0}, cosmology with a general stiff matter era was considered where for a stiff matter with negative energy density was shown to result in bouncing solutions of the universe as observed in loop quantum cosmology \cite{AshtekarSingh}.

\noindent Our aim and motivation in this paper is simple. We want to investigate the solution of renormalization flow improved Friedmann equations in the presence of a primordial stiff matter era. We also want to observe whether one can consider an anti-stiff matter component in the renormalization group improved framework of cosmology. We also want to investigate whether an inflation era is possible in such flow improved cosmology.

\noindent Our paper is organized as follows. In section(\ref{S2}), we discuss briefly the classical Friedmann equations and flow improved Friedmann equations. Then in section(\ref{S3}), we discuss the the modified equation of state using the relativistic thermodynamical approach. Next, in section(\ref{S4}) and section(\ref{S5}), we obtain the solutions of the flow improved Friedmann equations for different combinations of the various components present in the universe in the $k>m_{Pl}$ regime and in the $k<m_{Pl}$ regime or the perturbative regime. In section(\ref{S5B}), we have investigated whether it is possible to get bouncing solutions of the universe. Next, in section(\ref{S6}), we discuss inflation for a polytropic equation of state. Finally, in section(\ref{SConclusion}), we summarize our results.
\section{Brief review of the background model}\label{S2}
\subsection{Renormalization group improved cosmology and the Friedmann equations}\label{S2.A}
\noindent We start our discussion by considering the Friedmann-Lema\^{i}tre-Robertson-Walker metric or the FLRW metric for an isotropic and homogeneous model of the universe which is given by\footnote{The speed of light is taken to be unity.}
\begin{equation}\label{2.1}
ds^2=-dt^2+a^2(t)\left(\frac{dr^2}{1-\kappa r^2}+r^2d\theta^2+r^2\sin^2\theta+d\phi^2\right)
\end{equation}
where $a(t)$ denotes the scale factor of the universe and $\kappa=\{-1,0,1\}$. If one can consider that the universe is filled with a perfect fluid with energy density $\varepsilon(t)$ and isotropic pressure $P(t)$, then the corresponding energy-momentum tensor reads
\begin{equation}\label{2.2}
T_{\mu\nu}=(\varepsilon+P)u_\mu u_\nu+Pg_{\mu\nu}
\end{equation}
with $u_{\mu}=\{1,0,0,0\}$ in a co-moving frame of reference. %The Einstein's field equations with non-vanishing cosmological constant reads
%\begin{equation}\label{2.3}
%\begin{split}
%R_{\mu\nu}-\frac{1}{2}g_{\mu\nu}R=8\pi G T_{\mu\nu}-\Lambda g_{\mu\nu}~.
%\end{split}
%\end{equation}
%Setting $\mu=0$ and $\nu=0$ in eq.(\ref{2.3}), we arrive at the first Friedmann equation which is given by
%\begin{equation}\label{2.4}
%\left(\frac{\dot{a}}{a}\right)^2=\frac{8\pi G\varepsilon}{3}+\frac{\Lambda}{3}-\frac{3\kappa}{a^2}~.
%\end{equation}
%Similarly using the $\{r,r\}$ component from eq.(\ref{2.3}) and making use of eq.(\ref{2.4}), we obtain the second Friedmann equation as\footnote{The $\{\theta,\theta\}$ and $\{\phi,\phi\}$ components give the same analytical form as eq.(\ref{2.5}).}
%\begin{equation}\label{2.5}
%\frac{\ddot{a}}{a}=-\frac{4\pi G}{3}(\varepsilon+3P)+\frac{\Lambda}{3}+\frac{\kappa}{a^2}~.
%\end{equation}
%Again $\nabla^\mu G_{\mu\nu}=0$, which along with the metric compatibility condition gives from eq.(\ref{2.3})
%\begin{equation}\label{2.6}
%\nabla^\mu T_{\mu\nu}=0~.
%\end{equation} 
%Using eq.(\ref{2.2}) in the above equation, we obtain
%\begin{equation}\label{2.7}
%\frac{d\varepsilon}{dt}+\frac{3\dot{a}}{a}(\varepsilon+P)=0~.
%\end{equation}
As experimental evidence shows, our universe is almost flat. Hence, without loss of any generality, we can proceed with $\kappa=0$.
%, and write down the modified Friedmann equations as
%\begin{equation}\label{2.8}
%\begin{split}
%H^2&=\frac{8\pi G\varepsilon}{3}+\frac{\Lambda}{3}\\
%\frac{\ddot{a}}{a}&=-\frac{4\pi G}{3}(\varepsilon+3P)+\frac{\Lambda}{3}\\
%\frac{d\varepsilon}{dt}&=-\frac{3\dot{a}}{a}(\varepsilon+P)
%\end{split}
%\end{equation}
%where the Hubble constant is defined by $H\equiv\frac{\dot{a}}{a}$.
As our main focus is to investigate a stiff-matter dominated cosmology where renormalization group flows plays an important role, we shall start by briefly discussing the renormalization group approach to quantum gravity and cosmology developed in \cite{BonanoReuter,BonanoReuter2}. The primary way to such an approach is to consider an effective average action $\Gamma_k[g_{\mu\nu}]$ which depends on the flow of the momentum scale $k$. As a result of this renormalization group improved approach, the Newtonian gravitational constant $G$ and the cosmological constant $\Lambda$, flows with the momentum scale $k$ and can be written respectively as $G(k)$ and $\Lambda(k)$. In \cite{BonanoReuter,Reuter,Souma}, the differential equations governing the dynamics of $G(k)$ and $\Lambda(k)$ were obtained. As the momentum  scale $k$ approaches infinity, $G(k)$ approaches zero which implies that gravity is asymptotically free for such a theory. As $k\rightarrow\infty$, the dimensionless Newtonian constant $g(k)$, where $g(k)=k^2G(k)$, approaches an UV-attractive fixed point $g_*^{UV}$. As a result one can express the flowing gravitational constant for very high values of $k$ as \cite{Reuter}
\begin{equation}\label{2.9}
G(k)=\frac{g_*^{UV}}{k^2}~.
\end{equation}
The cosmological constant $\Lambda(k)$ for very high values of $k$ with an UV-attractive fixed point $\lambda_*^{UV}$ reads \cite{Reuter}
\begin{equation}\label{2.10}
\Lambda(k)=\lambda_*^{UV} k^2~.
\end{equation}
Here, $G(k)$ and $\Lambda(k)$ follow the asymptotic solutions given in eq.(s)(\ref{2.9},\ref{2.10}), if and only if $k$ is larger than the Planck mass \cite{BonanoReuter2}. The renormalization group improved Friedmann equations read \cite{BonanoReuter2} (for $\kappa=0$)
\begin{align}
\left(\frac{\dot{a}}{a}\right)^2&=\frac{8\pi G(k)\varepsilon}{3}+\frac{\Lambda(k)}{3}~,\label{2.11}\\
\frac{\ddot{a}}{a}&=-\frac{4\pi G(k)}{3}(\varepsilon+3P)+\frac{\Lambda(k)}{3}\label{2.12}
\end{align}
where $k\equiv k(t)$ is a function of $t$ as will be identified in the subsequenct discussion.
Doing a time derivative of eq.(\ref{2.11}) and making use of eq.(\ref{2.12}), we obtain 
\begin{equation}\label{2.13}
\frac{d\varepsilon}{dt}+\frac{3\dot{a}}{a}(\varepsilon+P)+\frac{\dot{\Lambda}(k(t))+8\pi \varepsilon\dot{G}(k(t))}{8\pi G(k(t))}=0~.
\end{equation}
%\textcolor{blue}{One can also make use of the Bianchi identity \cite{BonanoSaueressig} to obtain the above equation. We now argue that it is also important that the geometry part and matter part is covariantly conserved corresponding to the Einstein equations which leads us to implement
%the covariant conservation of stress energy tensor which is given by the condition $\nabla^\mu T_{\mu\nu}=0$. This is not a restriction rather a requirement when the geometry part and matter part is treated separately. Implementing this condition,} 
Implementing the covariant conservation of stress energy tensor which is given by the condition $\nabla^\mu T_{\mu\nu}=0$, we get back the standard continuity equation given as 
\begin{equation}\label{2.13a}
\frac{d\varepsilon}{dt}+\frac{3\dot{a}}{a}(\varepsilon+P)=0~.
\end{equation}
 In the presence of the flow of $G$ and $\Lambda$ with the momentum scale $k$. For a homogeneous and isotropic universe the scale factor $k$ can be a function of the cosmological time only \cite{BonanoReuter2}. Comparing eq.(\ref{2.13}) with the standard continuity equation in Friedmann-Lema\^{i}tre-Robertson-Walker cosmology, we obtain a constraint condition involving $\Lambda$, $G$, and $\varepsilon$ as  \cite{RGCosmo2}
\begin{equation}\label{2.14}
\begin{split}
&\dot{\Lambda}+8\pi \varepsilon \dot{G}=0\\
\implies&\varepsilon=-\frac{\dot{\Lambda}}{8\pi \dot{G}}=-\frac{d_k\Lambda(k)}{8\pi d_kG(k)}
\end{split}
\end{equation}
with the definition $d_k\equiv \frac{d}{dk}$. When $k\rightarrow\infty$ or at very early time ($t\ll t_{\text{Pl}}$), we can make use of eq.(\ref{2.9}) and eq.(\ref{2.10}) to obtain the energy density as a function of $k$ in the following form
\begin{equation}\label{2.15}
\varepsilon=\frac{\lambda_{*}^{UV}}{8\pi g_{*}^{UV}}k^4~.
\end{equation}
This indeed is a very important relation in a sense that one can express the energy-density of the universe in terms of the momentum scale introducing a flow-like condition of the energy density itself. It is imortant to note that if $G(k)$ and $\Lambda(k)$ are considered as time-dependent constants then eq.(\ref{2.13}) reduces to the standard continuity equation in eq.(\ref{2.13a}) turning eq.(\ref{2.14}) into a ``0=0" identity.
If $k\rightarrow0$ (for $t\gg t_{\text{Pl}}$), the dynamical relations  of $G$ and $\Lambda$ follows the following equations \cite{BonanoReuter2}
\begin{align}
G(k)&=G_0\left(1-\omega G_0 k^2+\mathcal{O}(G_0^2k^4)\right)\label{2.16}\\
\Lambda(k)&=\Lambda_0+\nu G_0k^4(1+\mathcal{O}(G_0k^2))\label{2.17}
\end{align}
where the analytical forms of $\omega$ and $\nu$ are given by
\begin{equation}\label{2.18}
\omega=\frac{1}{6\pi}\left(24\Phi_2^2(0)-\Phi_1^1(0)\right)~,~~\nu=\frac{1}{4\pi}\Phi_2^1(0)
\end{equation}
with the threshold function $\Phi_n^m(w)$ defined as \cite{BonanoReuter2} ($m=1,2,\ldots$)
\begin{equation}\label{2.19}
\Phi_n^m(w)=\frac{1}{\Gamma(n)}\int_0^\infty dz~z^{n-1}\frac{\left(R^{(0)}(z)-zR^{(0)'}(z)\right)}{\left(z+R^{(0)}(z)+w\right)^p}
\end{equation}
with $R^{(0)}$ denoting the cut-off function. In eq.(s)(\ref{2.16},\ref{2.17}), $G_0$ and $\Lambda_0$ denotes the value of the Newton's gravitational constant and cosmological constant in the present time. Using eq.(\ref{2.14}), and eq.(s)(\ref{2.16},\ref{2.17}), one can obtain the form of the energy density in terms of $k$ as (upto lowest order contribution in $k$) \cite{Cosmologyfinal}
\begin{equation}\label{2.20}
\varepsilon\simeq\frac{\nu}{4\pi \omega G_0}k^2~.
\end{equation}
Our primary aim in this work is to make use of the renormalization group improved Friedmann equations to investigate the implications of the existence of a stiff-matter era in the early universe and analyse its underlying implications.\footnote{This section is purely intended for a brief pedagogical review of the background model consisting of results from previous works that we shall use for our current analysis.}
\section{Modified equation of state using the relativistic thermodynamical approach}\label{S3}
The first law of thermodynamics reads
\begin{equation}\label{3.A.21}
\begin{split}
TdS&=dU+PdV\\
\implies Td(sV)&=d(\varepsilon V)+PdV
\end{split}
\end{equation}
where $s$ denotes the entropy density and $\varepsilon$ denotes the energy density for a system. If $N$ is the number of particles inside a box of volume $V$ with each particles having a mass $m$, then dividing the left hand side and right hand side of the above equation with $Nm$, we obtain the local form of the first law of thermodynamics \cite{Weinberg,Chavanis0}
\begin{equation}\label{3.A.22}
\begin{split}
Td\left(\frac{s}{nm}\right)&=d\left(\frac{\varepsilon}{nm}\right)+Pd\left(\frac{1}{n m}\right)\\
\implies Td\left(\frac{s}{\rho}\right)&=d\left(\frac{\varepsilon}{\rho}\right)+Pd\left(\frac{1}{\rho}\right)
\end{split}
\end{equation}
where $n=\frac{N}{V}$ denotes the number density and $\rho=nm$ is the mass density. Considering the fluid to be at zero temperature, one can obtain from eq.(\ref{3.A.22})
\begin{equation}\label{3.A.23}
\frac{d\varepsilon}{d\rho}=\frac{1}{\rho}(\varepsilon+P)~.
\end{equation}
Making use of the above equation in the standard continuity equation in FLRW cosmology, we obtain the following relation \cite{Chavanis0}
\begin{equation}\label{3.A.24}
\frac{d\rho}{dt}=-\frac{3\dot{a}}{a}\rho~.
\end{equation} 
Solving the above equation, one can obtain the following analytical form of the mass density in terms of the scale factor as
\begin{equation}\label{3.A.25}
\rho=\frac{\mathcal{A}}{a(t)^3}
\end{equation}
with $\mathcal{A}$ denoting the integration constant. Instead of fixing the constant using the current time values of $\rho$ and $a$, we focus mainly about the Planck time regime as we are mainly focused in investigating the dynamics of the universe before the Planck time ($t_{\text{Pl}}$). In the current analysis $t_*^{UV}=t_{\text{Pl}}$. At $t=t_*^{UV}$, $\rho=\rho_*^{UV}$ and $a=a_*^{UV}$\footnote{For removing cluttering in the expressions, we write denote the scale factor and the energy density of the universe at the fixed time as $a_*^{UV}=a_*$ and $\rho_{*}^{UV}=\rho_*$.}. Hence, we find that the integration constant has the value $\mathcal{A}={a_*^{UV}}^3\rho_*^{UV}=a_*^3\rho_*$. We can now recast eq.(\ref{3.A.25}) as
\begin{equation}\label{3.A.26}
\rho(t)=\rho_*\left(\frac{a_*}{a(t)}\right)^3.
\end{equation}
Within the fixed time regime (considering the universe is always expanding) ($t\ll t_{\text{Pl}}$), $a\ll a_*$ but for our calculation we shall make the range of the scale factor $a(t)$ a bit flexible by considering eq.(s)(\ref{2.9},\ref{2.10}) to hold when $a<a_*$ as well. Eq.(\ref{3.A.26}) is different from the usual structure as the reference point is the fixed-point regime instead of the current-time as is usually done in the literature \cite{Weinberg,
WeinbergCosmo,Ryden} .

\noindent Again from eq.(\ref{3.A.23}), we know that $P=P(\rho)$ and we obtain the analytical form of the energy density as a function of pressure and rest-mass density of the fluid as \cite{Chavanis0}
\begin{equation}\label{3.A.27}
\varepsilon(\rho)=\rho+u(\rho)
\end{equation}
where the analytical form of $u(\rho)$ is defined as
\begin{equation}\label{3.A.28}
u(\rho)\equiv\rho\int_{0}^{\rho}d\rho'\frac{P(\rho')}{{\rho'}^2}~.
\end{equation}
One can consider the equation of state of the fluid of the form
\begin{equation}\label{3.A.29}
P(\rho)=K_{\zeta}\rho^\zeta
\end{equation}
where $\zeta>1$. The above equation of state gives the analytical form of $u(\rho)$ to be $\frac{K_\zeta \rho^{\zeta}}{\zeta-1}$. Considering the universe to be filled with a dark fluid (described by a quadratic equation of state) and, we can express the energy density in eq.(\ref{3.A.27}) as
\begin{equation}\label{3.A.30}
\varepsilon(\rho)=\rho+K_2\rho^2~.
\end{equation}
According to the model proposed by Zel'dovich \cite{Zeldovich,Zeldovich2}, the complete equation of state describing the cosmological evolution process reads
\begin{equation}\label{3.A.30I}
P=K_2\rho^2
\end{equation}
and the energy density was expressed as
$\varepsilon(\rho)=\rho+K_2\rho^2$ which same as eq.(\ref{3.A.30}). The above expression for the energy density can be further simplified to the form given by\footnote{For reference see \cite{Chavanis0}.}
\begin{equation}\label{3.A.31}
\varepsilon(\rho)=\rho+K_2\rho^2+3K_{\frac{4}{3}}\rho^{\frac{4}{3}}~.
\end{equation}
Substituting $\rho$ from eq.(\ref{3.A.26}) in the above equation, we obtain
\begin{equation}\label{3.A.32}
\varepsilon(\rho)=\rho_*\left(\frac{a_*}{a}\right)^3+K_2\rho_*^2\left(\frac{a_*}{a}\right)^6+3K_{\frac{4}{3}}\rho_*^{\frac{4}{3}}\left(\frac{a_*}{a}\right)^4~.
\end{equation}
The above energy density when substituted in the Friedmann equation (eq.(\ref{2.11})), gives a universe with a matter component ($\varepsilon\propto a^{-3}$), a radiation ($\varepsilon\propto a^{-4}$), and a stiff matter component ($\varepsilon\propto a^{-6}$). A simplified way of writing the energy density as
\begin{equation}\label{3.A.33}
\varepsilon=\frac{\varepsilon_{\text{s},*}}{\left(\nicefrac{a}{a_*}\right)^6}+\frac{\varepsilon_{\text{rad},*}}{\left(\nicefrac{a}{a_*}\right)^4}+\frac{\varepsilon_{\text{m},*}}{\left(\nicefrac{a}{a_*}\right)^3}
\end{equation}
where $\varepsilon_{\text{m},*}$, $\varepsilon_{\text{rad},*}$, and $\varepsilon_{\text{s},*}$ denotes the energy densities corresponding to the matter part, radiation part and stiff matter part of the universe respectively. One can now write down the energy densities corresponding to each type of matter as
\begin{equation}\label{3.A.34}
\varepsilon_{\text{m},*}=\Omega_{\text{m},*}\varepsilon_*,~ \varepsilon_{\text{rad},*}=\Omega_{\text{rad},*}\varepsilon_*,~\text{and}~\varepsilon_{\text{s},*}=\Omega_{\text{s},*}\varepsilon_*
\end{equation}
where $\Omega_{\text{m},*},\Omega_{\text{rad},*}$, and $\Omega_{\text{s},*}$ are greater than or equal to zero for positive energy density. Using eq.(\ref{3.A.34}), we can recast eq.(\ref{3.A.33}) as
\begin{equation}\label{3.A.35}
\frac{\varepsilon}{\varepsilon_*}=\frac{\Omega_{\text{s},*}}{\left(\nicefrac{a}{a_*}\right)^6}
+\frac{\Omega_{\text{rad},*}}{\left(\nicefrac{a}{a_*}\right)^4}+\frac{\Omega_{\text{m},*}}{\left(\nicefrac{a}{a_*}\right)^3}
\end{equation}
which also implies that $\Omega_{\text{s},*}+\Omega_{\text{rad},*}+\Omega_{\text{m},*}=1$ as $\varepsilon=\varepsilon_*$ when $a=a_*$ (for $t=t_{\text{Pl}}$). 
The above energy density although have a structure similar to the one discussed in \cite{Chavanis0}, it uses Planck time ($t_*\equiv t_{Pl}$) as the time of reference whereas for the case in \cite{Chavanis0}, the reference time is the current time of our universe ($t_0$)\footnote{We have discussed the analytical form of the energy density (used in this analysis) in details to truly highlight the time of reference used here in contrast to the earlier analyses \cite{Chavanis0,Chavanis1,Chavanis2}.}.
\section{RG flow improved Friedmann equations with a stiff matter era above the Planck scale ($k>m_{Pl}$)}\label{S4}
\noindent In this section we shall investigate the renormalization flow improved Friedmann equations when there is a stiff matter era before the Planck time and we shall also discuss the evolution of the scale factor after the Planck time. Substituting $G(k)$, $\Lambda(k)$, and $\varepsilon(k)$ from eq.(s)(\ref{2.9},\ref{2.10},\ref{2.15}), in eq.(\ref{2.11}), we obtain the Friedmann equation in terms of $k$ as
\begin{equation}\label{4.36}
H^2=\frac{2\lambda_*^{UV} k^2}{3}=\frac{2\lambda_* k^2}{3}
\end{equation}
where we have made use of the identification $\lambda_*\equiv\lambda_*^{UV}$. From eq.(\ref{4.36}), we can see that $H=\frac{\dot{a}}{a}$ is dependent upon $k$. This dependence upon $k$ originates due to the flow of the cosmological constant $\Lambda$. Therefore, $H$ has no constant contribution.
\subsection{Stiff matter and radiation era}\label{S4.A}
\begin{figure}
\begin{center}
\includegraphics[scale=0.48]{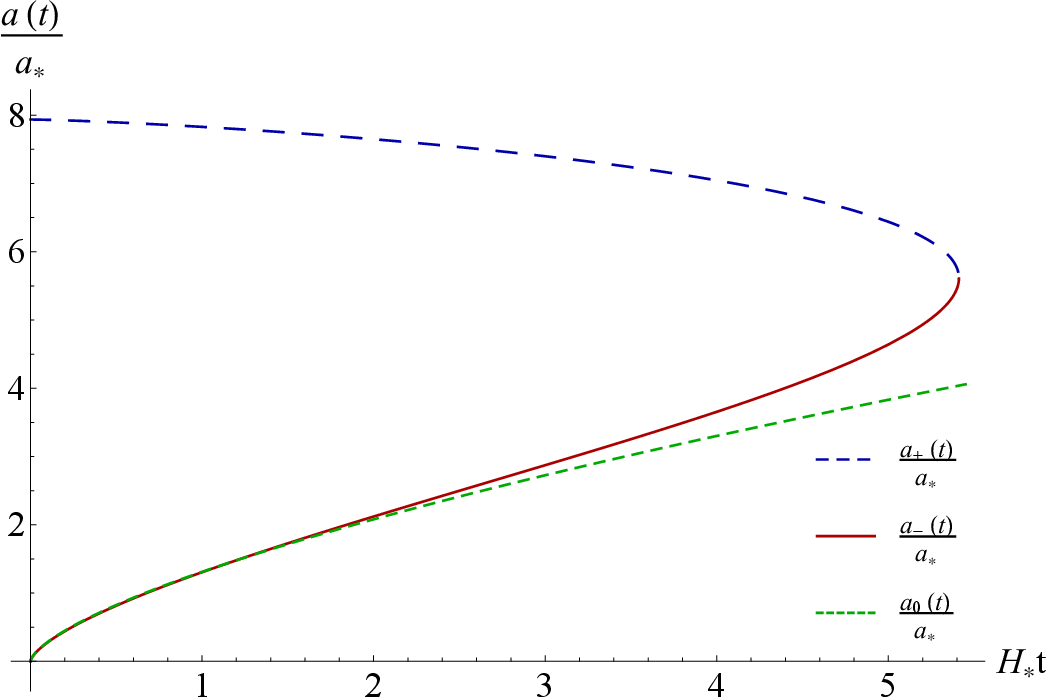}
\caption{Plot of $\frac{a_0(t)}{a_*}$, $\frac{a_+(t)}{a_*}$, and $\frac{a_-(t)}{a_*}$ versus the dimensionless time $H_*t$.\label{Fig1}}
\end{center}
\end{figure}

\begin{figure}
\begin{center}
\includegraphics[scale=0.48]{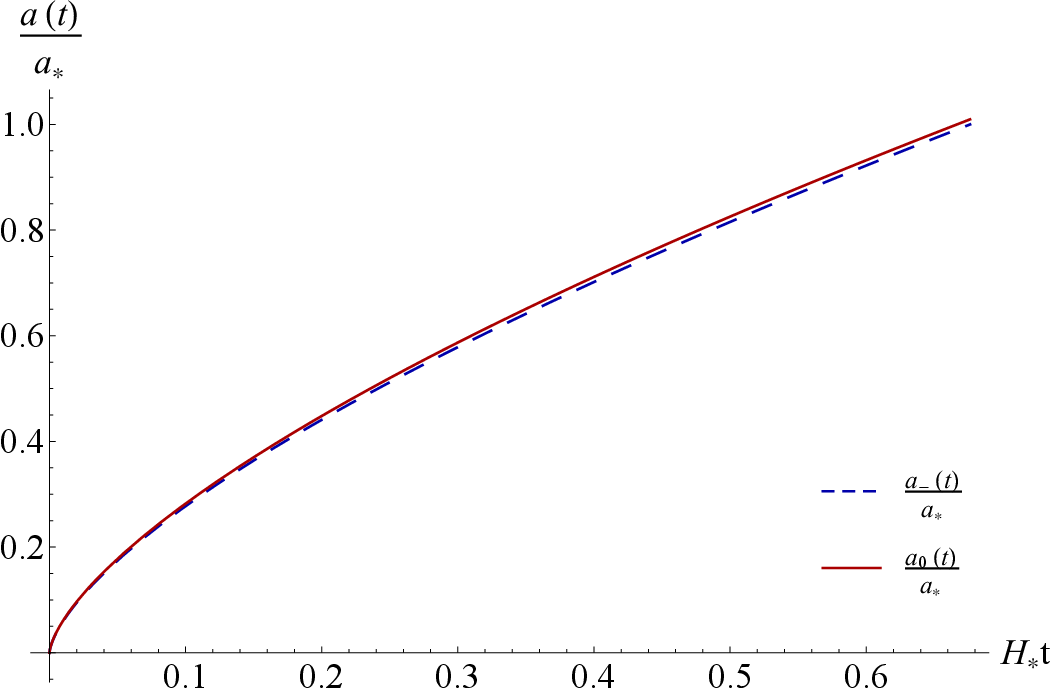}
\caption{Plot of $\frac{a_0(t)}{a_*}$ and $\frac{a_-(t)}{a_*}$ versus the dimensionless time $H_*t$. Here is plot is terminated at $H_*t=H_*t_*$\label{Fig1a}.}
\end{center}
\end{figure}

\begin{figure}
\begin{center}
\includegraphics[scale=0.48]{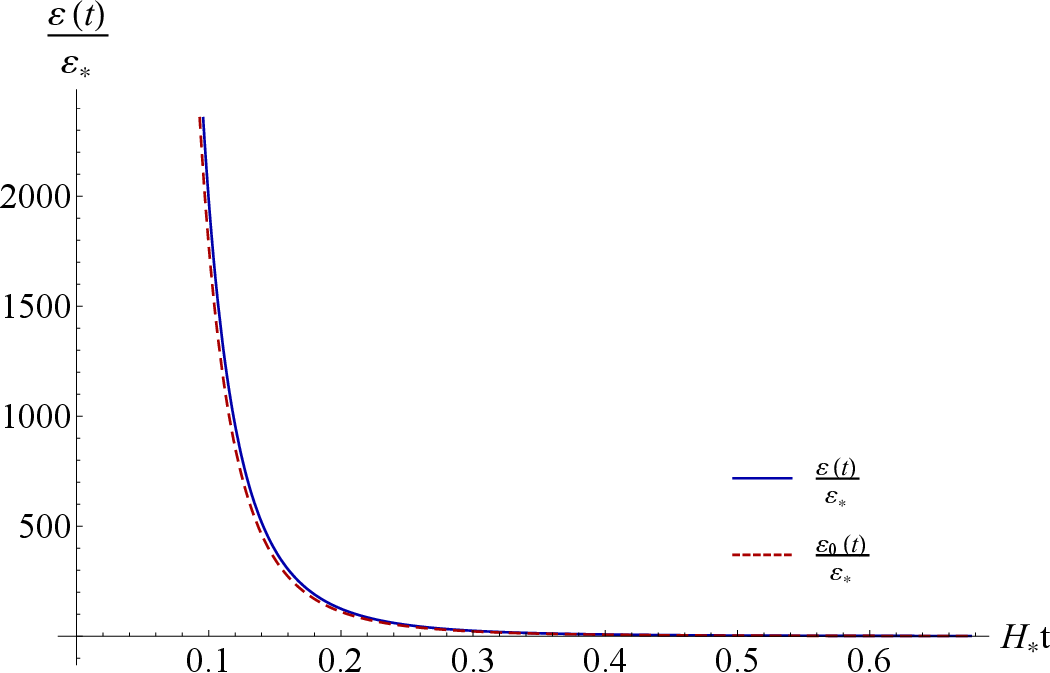}
\caption{Plot of $\frac{\varepsilon(t)}{\varepsilon_*}$ against $H_* t$ for the cases when radiation dominated early era is both present and absent ($\varepsilon_0(t)$ denotes universe in the absence of any early radiation dominated era). \label{Fig2}}
\end{center}
\end{figure}

\noindent Just after the inflationary phase, the universe enters into the reheating epoch. This reheating era heats up the cold and inflated universe. It is therefore largely considered that the radiation era existed just after the inflation era near to the beginning of nucleosynthesis. The inflation era is assumed to happen between $10^{-33}$ sec and $10^{-32}$ sec which is way after the Planck time of the universe. It was assumed in \cite{Zeldovich,Zeldovich2}, that the universe consisted of a cold baryon gas which obeys a stiff equation of state and contributed to the evolution of the universe in the post-inflation period and just before the radiation era. In \cite{OdintsovOikonomou}, a stiff matter era was considered in the post-inflationary period in the case of $f(R)$ gravity. In case of going with the traditional way of incorporating stiff-matter era, we assume the presence of a dark-fluid which behaves as a stiff-matter fluid inside the scaling regime when $t<t_{Pl}$. This allows one for a very early ``power-law" expansion of the universe.  In the cosmological renormalization group improved models, before the Planck time, the energy density is very high and as a result it is quite clearly understandable that any fluid existing at such an early time will have pressure so high that the speed of sound will approach the speed of light. As argued in \cite{Chavanis0}, the universe at earlier times is considered to be filled with a dark-fluid. In the $0<t\leq t_{Pl}$, we consider the energy density of the dark-fluid to have a $a^{-6}$ dependence and a $a^{-4}$ dependence. We incorporate a radiation-like part in the energy density which is expected to dominate after the Plack time. In contrast to the general cosmological models here, one considers a single fluid which shows different behaviour with respect to the change in time. 
We shall now therefore consider the universe with an early stiff matter and radiation era. For such a model, the energy density of the universe varies with the scale factor as
\begin{equation}\label{4.A.37}
\frac{\varepsilon}{\varepsilon_*}=\frac{\Omega_{\text{s},*}}{\left(\nicefrac{a}{a_*}\right)^6}+\frac{\Omega_{\text{rad},*}}{\left(\nicefrac{a}{a_*}\right)^4}~.
\end{equation}
From the above equation, we can find out that when the scale factor is zero, the energy density goes to infinity and when $a\rightarrow a_*$, $\varepsilon\rightarrow\varepsilon_*$ as $\Omega_{\text{rad},*}+\Omega_{\text{s},*}=1$ in this case. 
Comparing eq.(\ref{2.15}) with eq.(\ref{4.A.37}), we can express the momentum scale $k$ in terms of the scale factor of the universe as
\begin{equation}\label{4.A.38}
k=\left(\frac{8\pi g_*\varepsilon_*}{\lambda_*}\right)^{\frac{1}{4}}\left(\frac{\Omega_{\text{rad},*}}{\left(\nicefrac{a}{a_*}\right)^4}+\frac{\Omega_{\text{s},*}}{\left(\nicefrac{a}{a_*}\right)^6}\right)^{\frac{1}{4}}~.
\end{equation}
Substituting the above equation in eq.(\ref{4.36}), we obtain
\begin{equation}\label{4.A.39}
H^2=\frac{2}{3}(8\pi g_*\lambda_*\varepsilon_*)^{\frac{1}{2}}\left(\frac{\Omega_{\text{rad},*}}{\left(\nicefrac{a}{a_*}\right)^4}+\frac{\Omega_{\text{s},*}}{\left(\nicefrac{a}{a_*}\right)^6}\right)^{\frac{1}{2}}~.
\end{equation}
We can define a new quantity $H_*$ which is the Hubble parameter at the Planck time by substituting $a_*$ in place of $a$  and obtain
\begin{equation}\label{4.A.40}
H_*^2=\frac{2}{3}(8\pi g_*\lambda_*\varepsilon_*)^{\frac{1}{2}}~.
\end{equation} 
Using the above equation back in eq.(\ref{4.A.39}), we obtain
\begin{equation}\label{4.A.41}
\frac{H^2}{H_*^2}=\left(\frac{\Omega_{\text{rad},*}}{\left(\nicefrac{a}{a_*}\right)^4}+\frac{\Omega_{\text{s},*}}{\left(\nicefrac{a}{a_*}\right)^6}\right)^{\frac{1}{2}}~.
\end{equation}
If we define a new variable $\xi\equiv\frac{a}{a_*}$, then the minimum value that $\xi$ can attain is zero and the upper limit will always be less than unity as inside the fixed point scaling regime $a<a_*$. Our primary aim now is to obtain the time evolution of the scale factor $a(t)$.
In terms of the new variable $\xi$, we can express $H$ as
\begin{equation}\label{4.A.42}
\begin{split}
H&=\frac{\dot{a}}{a}=\frac{\left(\nicefrac{\dot{a}}{a_*}\right)}{\left(\nicefrac{a}{a_*}\right)}\\
\implies H&=\frac{\dot{\xi}}{\xi}~.
\end{split}
\end{equation}
Taking a square-root and integrating both sides of eq.(\ref{4.A.42}), we obtain
\begin{equation}\label{4.A.43}
\begin{split}
H_*t=&\int_0^\xi\frac{d\xi~\xi^{\frac{1}{2}}}{\left(\xi^2 \Omega_{\text{rad},*}+\Omega_{\text{s},*}\right)^{\frac{1}{4}}}\\
=&\frac{2\xi^{\frac{3}{2}}}{3\Omega_s}\left(\Omega_{\text{rad},*}\xi^2+\Omega_{\text{s},*}\right)^\frac{3}{4}\tensor[_{2}]{F}{_{1}}\left[1,\frac{3}{2},\frac{7}{4},-\frac{\Omega_{\text{rad,*}}\xi^2}{\Omega_{\text{s},*}}\right].
\end{split}
\end{equation}
In the $t<t_* (t_{Pl})$ regime, $\xi<1$, and as a result it is always quite straightforward to obtain a series expansion of the Gauss hypergeometric function in the above equation considering $\xi$ to be the small parameter. The problem with the exact solution is that it is impossible to obtain $\xi$ in terms of the time $t$. However, upto leading order contribution, we obtain $a$ in terms of $t$ as
\begin{equation}\label{4.A.44}
a(t)\simeq a_*\left(\frac{3\Omega_{\text{s},*}^{\frac{1}{4}}H_* t}{2}\right)^{\frac{2}{3}}~.
\end{equation} 
We find out that in the fixed point scaling regime the leading order contribution to the evolution of the scale factor will come from the stiff-matter fluid with a time dependece proportional to $t^{\frac{2}{3}}$. In order to obtain the contribution due to the radiation term, one needs to consider higher order contributions in $\xi$ in eq.(\ref{4.A.43}). Expanding the right hand side of eq.(\ref{4.A.43}) in powers of $\xi$, we can express eq.(\ref{4.A.43}) as
\begin{equation}\label{4.A.45}
\begin{split}
H_*t\simeq \frac{2\xi^{\frac{3}{2}}}{3\Omega_{\text{s},*}^{\frac{1}{4}}}\left(1-\frac{3\Omega_{\text{rad},*}\xi^2}{28\Omega_{\text{s},*}}+\mathcal{O}(\xi^4)\right)~.
\end{split}
\end{equation}
Taking a $\frac{4}{3}$ power of the both sides of the above equation, we obtain
\begin{equation}\label{4.A.46}
\begin{split}
\left(\frac{3\Omega_{\text{s},*}H_*t}{2}\right)^\frac{4}{3}&=\Omega_{\text{s},*}\xi^2\left(1-\frac{\Omega_{\text{rad},*}\xi^2}{7\Omega_{\text{s},*}}\right)\\
&=\Omega_{\text{s},*}\xi^2-\frac{\Omega_{\text{rad},*}\xi^4}{7}\\
\implies 0&=\xi^4-\frac{7\Omega_{s,*}}{\Omega_{\text{rad},*}}\xi^2+\frac{7}{\Omega_{\text{rad},*}}\left(\frac{3\Omega_{\text{s},*}H_*t}{2}\right)^\frac{4}{3}~.
\end{split}
\end{equation}
Solving, the above equation, we obtain the solution of the scale factor as a function of time as
\begin{equation}\label{4.A.47}
\begin{split}
\frac{a_{\pm}(t)}{a_*}\simeq \left[\frac{7\Omega_{\text{s},*}}{2\Omega_{\text{rad},*}}\pm \sqrt{\frac{49\Omega^2_{\text{s},*}}{4\Omega^2_{\text{rad},*}}-\frac{7}{\Omega_{\text{rad},*}}\left(\frac{3\Omega_{\text{s},*}H_*t}{2}\right)^\frac{4}{3}}\right]^{\frac{1}{2}}.
\end{split}
\end{equation}
We now need to figure out whether the minus or the positive sign in the above solution leads to the correct nature of the scale factor and plot the two solutions against time. We shall now denote the leading order solution from eq.(\ref{4.A.44}) as $a_0(t)$ and the solutions containing higher order contributions in eq.(\ref{4.A.47}) are already termed as $a_{+}(t)$ and $a_{-}(t)$ corresponding to the sign after $\frac{7\Omega_{\text{s},*}}{2\Omega_{\text{rad},*}}$, and compare the behaviour of the dimensionless scale factor $\frac{a(t)}{a_*}$ with respect to the dimensionless time $H_*t$ in Fig.(\ref{Fig1}). For the parameters, we have set $\Omega_{\text{s},*}=0.9$ and $\Omega_{\text{rad},*}=0.1$. 
From Fig.(\ref{Fig1}), we observe that with increase in time $\frac{a_+(t)}{a_*}$ decreases which goes against the well-known behaviour of the universe leading to a phantom behaviour. If one plots the energy density from eq.(\ref{4.A.37}) using $a_{+}(t)$, one observes that the energy density increases with time which confirms this phantom-like behaviour for this particular time dependence of the scale factor. Whereas we can see that both $a_-(t)$ and $a_0(t)$ shows quite a convincing behaviour as with increase in time the scale factor increases. It is though important to understand that as $\Omega_{\text{rad},*}$ approaches zero, $a_{-}(t)\rightarrow a_0(t)$, which is an expected behaviour.  We therefore consider $a_-(t)$ to show the most convincing behaviour of the scale factor and set it as $a(t)$. In principle, $a_0(t)$ shows the time-dependence of the scale factor in absence of any radiation as can be obtained directly by setting $\Omega_{\text{rad},*}=0$ in eq.(\ref{4.A.43}) (In such a scenario $\Omega_{\text{s},*}=1$). Using $a_0(t)$ and $a_-(t)$ from eq.(s)(\ref{4.A.44},\ref{4.A.47}) in eq.(\ref{4.A.37}), we get the time depence of the energy density of the universe. The $a_0(t)$ denotes the energy density of the universe in absence of radiation defined as $\varepsilon_0(t)$, whereas $a(t)$ gives the energy density in presence of both stiff matter and radiation component before the fixed time. It is although important to note that in Fig.(\ref{Fig1}), $\frac{a}{a_*}$ has exceeded unity. From the $\frac{a}{a_*}=1$ condition, we can obtain the value of $H_*t$, which in the current case is obtained to be $H_*t\simeq0.68$. Hence, upto $\frac{a}{a_*}=1$, we can plot the scale factor versus the dimensionless times in Fig.(\ref{Fig1a}). 
We now plot the dimensionless energy density $\left(\frac{\varepsilon(t)}{\varepsilon_*}\right)$ against the dimensionless time ($H_*t$) in Fig.(\ref{Fig2}) for $\Omega_{\text{rad},*}=0.1$ and $\Omega_{\text{s},*}=0.9$.
We observe from Fig.(\ref{Fig2}), that with increase in time the energy density of the universe decreases which is an expected behaviour and both $\varepsilon_0(t)$ and $\varepsilon(t)$ shows almost similar behaviour proving  dominance of a stiff matter era in the very early universe. For the solution of $a(t)$ presented in eq.(\ref{4.A.47}), one obtain a restriction on upto which time the evolution of the scale factor is reasonable and the inequality is obtained by claiming that the term inside the square root must be greater  than or equals to zero. This argument gives us the following inequality on the time as
\begin{equation}\label{4.A.48}
\begin{split}
t\leq \frac{2}{3\Omega_{\text{s},*}H_*}\left(\frac{7\Omega_{\text{s},*}^2}{4\Omega_{\text{rad},*}}\right)^{\frac{3}{4}}~.
\end{split}
\end{equation}
Now the above inequality only points toward the validity of the solution presented in eq.(\ref{4.A.47}). 
Some significant changes of this flow improved cosmology with respect to the standard FLRW cosmology are in order. 
\begin{enumerate}
\item The dependence of the scale factor in the stiff-matter dominated era goes as $t^{\frac{2}{3}}$ for the renormalization group improved cosmological model whereas for the standard FLRW cosmology the scale factor has the time dependence given by $t^{\frac{1}{3}}$ \cite{Chavanis0}. This implies that for our case the expansion in the stiff-matter dominated era will be enhanced during the fixed point scaling regime which is the regime between the Big Bang and the Planck time.
\item Again, in the dimensionless time parameter $H_*t\equiv H(t_{Pl})t$, the Hubble's constant will be much higher as $H(t_{Pl})\gg H_0$ leading to a more enhanched ``power-law" expansion compared to the standard FLRW case inside the fixed point scaling regime before Planck time.
\item The flow improved cosmology portrays clearly that the domination of the stiff matter era ended when the relative scale factor becomes unity at the Planck time whereas for the FLRW cosmology the clear transition point between stiff-matter dominated era and radiation era is unclear \cite{Chavanis0}.
\end{enumerate}
In the next subsection, we shall also incorporate a dark-matter like part in the energy density and observe whether it has any substantial effect inside the fixed point scaling regime.
\subsection{Stiff matter, radiation, and dark matter}\label{S4.B}
\noindent We consider that the universe was filled with a dark fluid which has a stiff matter-like, radiation-like, and dark matter-like contributions and the corresponding energy density is given by eq.(\ref{3.A.35}). At the fixed point we get $\varepsilon\rightarrow\varepsilon_*$, which leads to the condition
\begin{equation}\label{4.B.49}
\Omega_{\text{s},*}+\Omega_{\text{rad},*}+\Omega_{\text{m},*}=1
\end{equation}
which has already been discussed after eq.(\ref{3.A.35}). As in has been done in the earlier subsection, we compare eq.(\ref{2.15}) with eq.(\ref{3.A.35}) and obtain $k$ in terms of the scale factor $a(t)$ as
\begin{equation}\label{4.B.50}
k=\left(\frac{8\pi g_*\varepsilon_*}{\lambda_*}\right)^{\frac{1}{4}}\left(\frac{\Omega_{\text{s},*}}{\left(\nicefrac{a}{a_*}\right)^6}+\frac{\Omega_{\text{rad},*}}{\left(\nicefrac{a}{a_*}\right)^4}+\frac{\Omega_{\text{m},*}}{\left(\nicefrac{a}{a_*}\right)^3}\right)^{\frac{1}{4}}~.
\end{equation}
Using the analytical form of $k$ in the first Friedmann equation given in eq.(\ref{4.36}), and making use of eq.(\ref{4.A.40}), we arrive at the Friedmann equation (when stiff matter, radiation, and dark matter is considered) as
\begin{equation}\label{4.B.51}
\frac{H^2}{H_*^2}=\left(\frac{\Omega_{\text{s},*}}{\left(\nicefrac{a}{a_*}\right)^6}+\frac{\Omega_{\text{rad},*}}{\left(\nicefrac{a}{a_*}\right)^4}+\frac{\Omega_{\text{m},*}}{\left(\nicefrac{a}{a_*}\right)^3}\right)^\frac{1}{2}~.
\end{equation}
One important thing to observe from the above equation is that the right hand side has no time independent constant terms and as a result, in this flow-improved cosmology, an early-time inflation is not possible. This conclusion is in line with the arguments presented in \cite{BonanoReuter2}. Inflation can be realized by introducing new equations of state which we shall present in section (\ref{S6}).

\begin{figure}
\begin{center}
\includegraphics[scale=0.48]{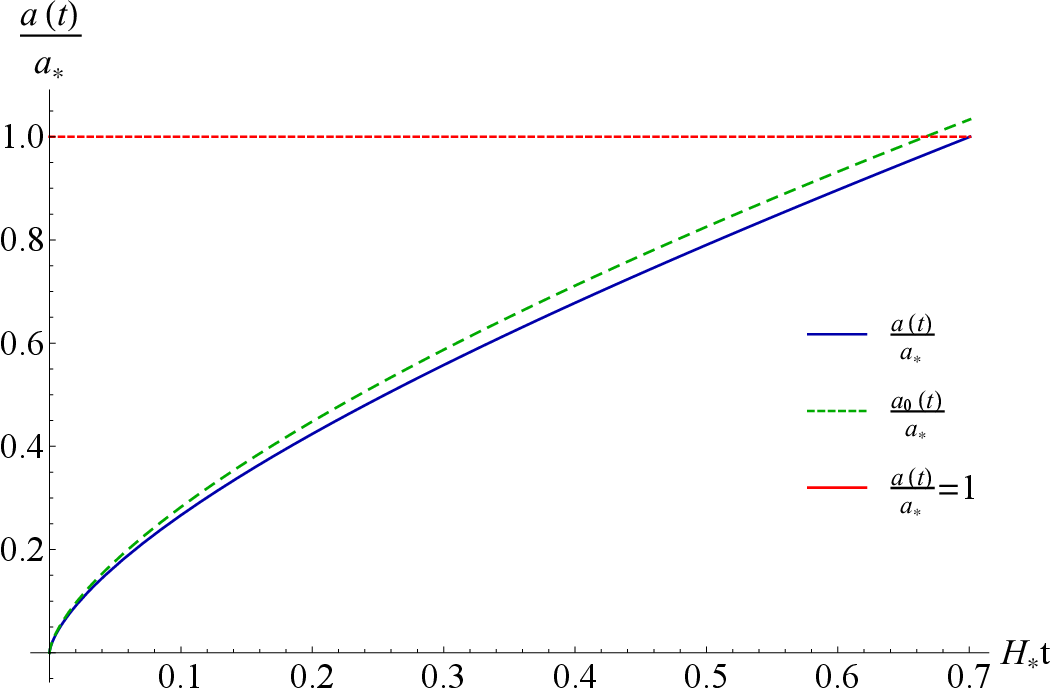}
\caption{Plot of $\frac{a(t)}{a_*}$ versus $H_*t$ for $\Omega_{\text{s},*}=0.7$, $\Omega_{\text{rad},*}=0.2$, and $\Omega_{\text{m},*}=0.1$ against the case when there is only stiff matter domination in the early universe ($\Omega_{\text{s},*}=1$).\label{Fig3}}
\end{center}
\end{figure}

\begin{figure}
\begin{center}
\includegraphics[scale=0.48]{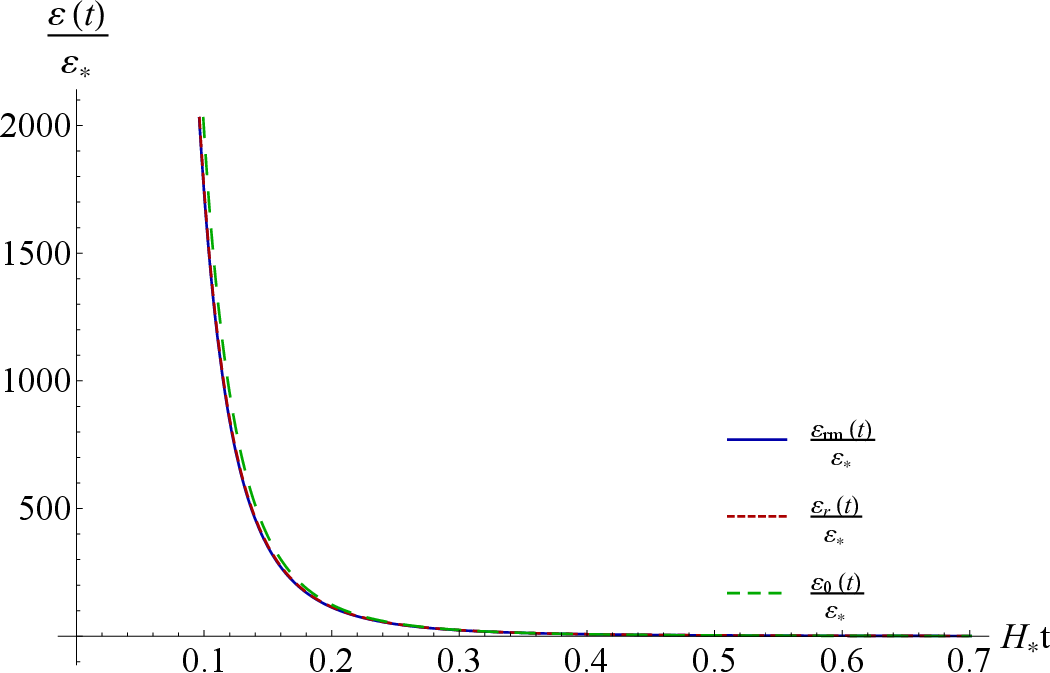}
\caption{Dimensionless energy density is plotted against dimensionless time for three cases when there is stiff-matter+radiation+dark-matter in the early universe, stiff-matter+radiation in the early universe, and only stiff-matter in the early universe.\label{Fig4}}
\end{center}
\end{figure}

\noindent Again substituting $\xi=\frac{a}{a_*}$, and carrying out the integration in the both sides of eq.(\ref{4.B.51}), we can recast the above expression as
\begin{equation}\label{4.B.52}
H_* t=\int_0^\xi\frac{d\xi~\xi^{\frac{1}{2}}}{\left(\xi^3\Omega_{\text{m},*}+\xi^2\Omega_{\text{rad},*}+\Omega_{\text{s},*}\right)^\frac{1}{4}}~.
\end{equation}
The above integration cannot be done exactly. Now, as the upper limit of integration is always less than unity we can do a Taylor series expansion of the term sitting in the denominator as
\begin{equation}\label{4.B.53}
\begin{split}
H_*t&\simeq\frac{1}{\Omega_{\text{s},*}^\frac{1}{4}}\int_0^\xi d\xi\left(\xi^\frac{1}{2}-\frac{\Omega_{\text{rad},*}\xi^\frac{5}{2}}{4\Omega_{\text{s},*}}-\frac{\Omega_{\text{m},*}\xi^\frac{7}{2}}{4\Omega_{\text{s},*}}\right)\\
&\simeq \frac{2\xi^\frac{3}{2}}{3\Omega_{\text{s},*}^{\frac{1}{4}}}\left(1-\frac{3\Omega_{\text{rad},*}\xi^{2}}{28\Omega_{\text{s},*}}-\frac{\Omega_{\text{m},*}\xi^{3}}{12\Omega_{\text{s},*}}\right)~.
\end{split}
\end{equation}
We cannot solve the above equation exactly just like we have done to obtain the time dependence of the scale factor in eq.(\ref{4.A.47}). Instead, we resort to a perturbative approach. We already have considered that the stiff matter era dominates in the early universe. Hence, there is no issue in considering $\Omega_{\text{s},*}>\Omega_{\text{rad},*},\Omega_{\text{m},*}$. It is also evident that the radiation part is more dominant in the energy density in eq.(\ref{3.A.35}) when $a<a_*$ than the dark matter part. As $\frac{\Omega_{\text{rad}*}}{\Omega_{\text{s},*}}<1$, we can always simplify eq.(\ref{4.A.47}) upto $\mathcal{O}\left(\frac{\Omega_{\text{rad}*}}{\Omega_{\text{s},*}}\right)$ as
\begin{equation}\label{4.B.54}
\begin{split}
\frac{a(t)}{a_*}\simeq \left(\frac{3\Omega_{\text{s},*}^{\frac{1}{4}}H_*t}{2}\right)^\frac{2}{3}+\frac{\Omega_{\text{rad},*}}{14\Omega_{\text{s},*}}\left(\frac{3\Omega_{\text{s},*}^{\frac{1}{4}}H_*t}{2}\right)^2~.
\end{split}
\end{equation}
We now propose a solution of the form given as
\begin{equation}\label{4.B.55}
\frac{a(t)}{a_*}\simeq \left(\frac{3\Omega_{\text{s},*}^{\frac{1}{4}}H_*t}{2}\right)^\frac{2}{3}+\frac{\Omega_{\text{rad},*}}{14\Omega_{\text{s},*}}\left(\frac{3\Omega_{\text{s},*}^{\frac{1}{4}}H_*t}{2}\right)^2+\frac{\zeta(t)\Omega_{\text{m},*}}{\Omega_{\text{s},*}}
\end{equation}
with $\zeta(t)$ being function of time which needs to be determined. Keeping terms upto $\mathcal{O}\left(\frac{\Omega_{\text{m}*}}{\Omega_{\text{s},*}}\right)$ and $\mathcal{O}\left(\frac{\Omega_{\text{rad}*}}{\Omega_{\text{s},*}}\right)$, and substituting the above solution back in eq.(\ref{4.B.53}), we obtain the analytical form of $\zeta(t)$ as
\begin{equation}\label{4.B.56}
\zeta(t)\simeq\frac{1}{18}\left(\frac{3\Omega_{\text{s},*}^{\frac{1}{4}}H_*t}{2}\right)^\frac{8}{3}.
\end{equation}
Substituting the analytical form of $\zeta(t)$ from the above equation in eq.(\ref{4.B.55}), we have
\begin{equation}\label{4.B.57}
\begin{split}
\frac{a(t)}{a_*}\simeq& \left(\frac{3\Omega_{\text{s},*}^{\frac{1}{4}}H_*t}{2}\right)^\frac{2}{3}+\frac{\Omega_{\text{rad},*}}{14\Omega_{\text{s},*}}\left(\frac{3\Omega_{\text{s},*}^{\frac{1}{4}}H_*t}{2}\right)^2\\&+\frac{\Omega_{\text{m},*}}{18\Omega_{\text{s},*}}\left(\frac{3\Omega_{\text{s},*}^{\frac{1}{4}}H_*t}{2}\right)^\frac{8}{3}~.
\end{split}
\end{equation}
We now plot the above result with respect to the case when there is no dark matter or radiation-like component of the dark fluid in the universe implying $\Omega_{\text{s},*}=1$. For the current case, we set, $\Omega_{\text{s},*}=0.7$, $\Omega_{\text{rad},*}=0.2$, and $\Omega_{\text{m},*}=0.1$. It is to be noted that the values that we have chosen here are in sharp contrast to the those that have been taken in \cite{Chavanis0}. In particular, we have considered $\Omega_{s,*}>\Omega_{\text{rad},*}>\Omega_{\text{m},*}$. The physical motivation behind this consideration lie in the fact that near the fixed point regime the dark-fluid primarily behaves as a stiff-matter whereas radiation and matter like components remain primarily latent.
We now plot the dimensionless scale factor against the dimensionless time in Fig.(\ref{Fig3}).
In the above plot, we observe that in the very early time stiff matter-like component dominates the evolution of the scale factor. In the later part, there is slight contribution due to radiation-like and dark matter-like component of the dark fluid. As expected, the dark-matter or radiation like component has almost no significant effect inside the fixed point regime implying that for $t<t_*$, the dark-fluid behaves primarily as a stiff matter, and dominates the expansion of the universe. Hence, without loss of any generality, one can consider eq.(\ref{3.A.35}) as the most general energy density for a dark-fluid in a renormalization flow improved cosmology. The rationale behind taking the other components, namely, the radiation and dark-matter is that these components will start to play an important role at later epochs as will be discussed subsequently. In Fig.(\ref{Fig3}), the dotted red line denotes the $\frac{a(t)}{a_*}=1$ line. We now plot the dimensionless energy densities for the three cases when there is stiff-matter+radiation+dark-matter, stiff-matter+radiation, and stiff-matter only in the universe against the dimensionless time. For the stiff-matter+radiation+dark-matter case, the energy density is described by $\varepsilon_{\text{rm}}(t)$, for the stiff-matter case the energy density of the universe is described by $\varepsilon_{\text{r}}(t)$, and for the stiff-matter only case, the energy density is described by $\varepsilon_0(t)$ in Fig.(\ref{Fig4}). In Fig.(\ref{Fig4}), for the $\varepsilon_{\text{rm}}$ case $\Omega_{\text{s},*}=0.6$, $\Omega_{\text{rad},*}=0.25$, and $\Omega_{\text{m},*}=0.15$, whereas for the $\varepsilon_{\text{r}}$ case, $\Omega_{\text{s},*}=0.6$ and $\Omega_{\text{rad},*}=0.4$.
From Fig.(\ref{Fig4}), we observe that the fall of the energy density with time is almost similar for all three cases, although the stiff-matter only case has a slower fall than the $\varepsilon_{\text{r}}(t)$ and $\varepsilon_{\text{rm}}(t)$ cases. 
\noindent One now look for other possible combinations. It is important to note that in any scenario, an anti-stiff matter era is impossible before the fixed point regime as it will indicate the existence of a zero energy-density which is in the contradiction with the renormalization group improved approach to cosmology. It is also evident that the coefficient of the lowest power of $a$ can not have negative values for a renormalization group improved cosmology. In such a scenario, only a perturbative analysis is sufficient and the concept of a UV-cut off becomes completely unecessary. This is in contrast to the anti-stiff matter scenario considered in \cite{Chavanis0}. The possibility of flow of the Newton's gravitational constant and the cosmological constant due to the renormalization group approach hinders the existence of the anti-stiff matter component in case of the fixed point analysis\footnote{We have investigated the case of a dark-fluid with negative energy density-components other than the anti-stiff matter in Appendix (\ref{A1}).}. It is though interesting whether bouncing solutions can exist. Hence, one can again point out the standard differences of this flow improved cosmological model with the standard FLRW cosmological model and some important aspects of this flow improved cosmological model as follows:
\begin{enumerate}
\item For a renormalization group improved cosmology a leading order negative energy density component cannot exist before the Planck time as it will lead to a singularity free zero-energy density region rulling out the need for ultra-violet fixed points. If such leading order contributions exist then it rules-out the need of any high energy analysis of the Friedmann equations and the perturbative corrections to the Friedmann equation becomes sufficient. 
\item Again it is evident that the stiff-matter era dominated before the Planck time and then the radiation and eventually matter era dominated around and after the Planck time.
\item As the cosmological constant flows with time, the Hubble parameter does not have any constant contributions which is again in contrary to FLRW cosmological model \cite{Chavanis0}.
\end{enumerate}
\section{Flow improved cosmology after the fixed point regime}\label{S5}
\noindent We shall now make use of eq.(s)(\ref{2.16},\ref{2.17}) in the Friedmann equations to investigate the behaviour of the scale factor and energy density after the fixed point regime. Using the forms of $G(k)$, $\Lambda(k)$, and $\varepsilon$ from eq.(s)(\ref{2.16},\ref{2.17},\ref{2.20}) and substituting it back in eq.(\ref{2.11}), we can obtain the Friedmann equation after the fixed point regime upto $\mathcal{O}(k^4)$ as
\begin{equation}\label{5.69}
H^2\simeq\frac{\Lambda_0}{3}+\frac{2\nu k^2}{3\omega}-\frac{\nu G_0 k^4}{3}~.
\end{equation}
Instead of now going back to the fixed point, we now set $k\rightarrow 0$ limit which implies the current time regime. In the $k\rightarrow 0$ limit
\begin{equation}\label{5.70}
\begin{split}
H^2\simeq\frac{\Lambda_0}{3}~.
\end{split}
\end{equation}
The above result leads to the solution
\begin{equation}\label{5.71}
a_0(t)\simeq e^{\sqrt{\frac{\Lambda_0}{3}}t}
\end{equation}
which indicate the accelerated expansion due to dark-energy in the current time. It is easy to identify from eq.(\ref{5.71}) that $H_0=\sqrt{\frac{\Lambda_0}{3}}$. Now we shall start with few of the cases that we have discussed in section (\ref{S4}). 
\subsection{Stiff matter, radiation, and dark matter era after the Planck time}\label{S5.A}
\begin{figure}
\begin{center}
\includegraphics[scale=0.48]{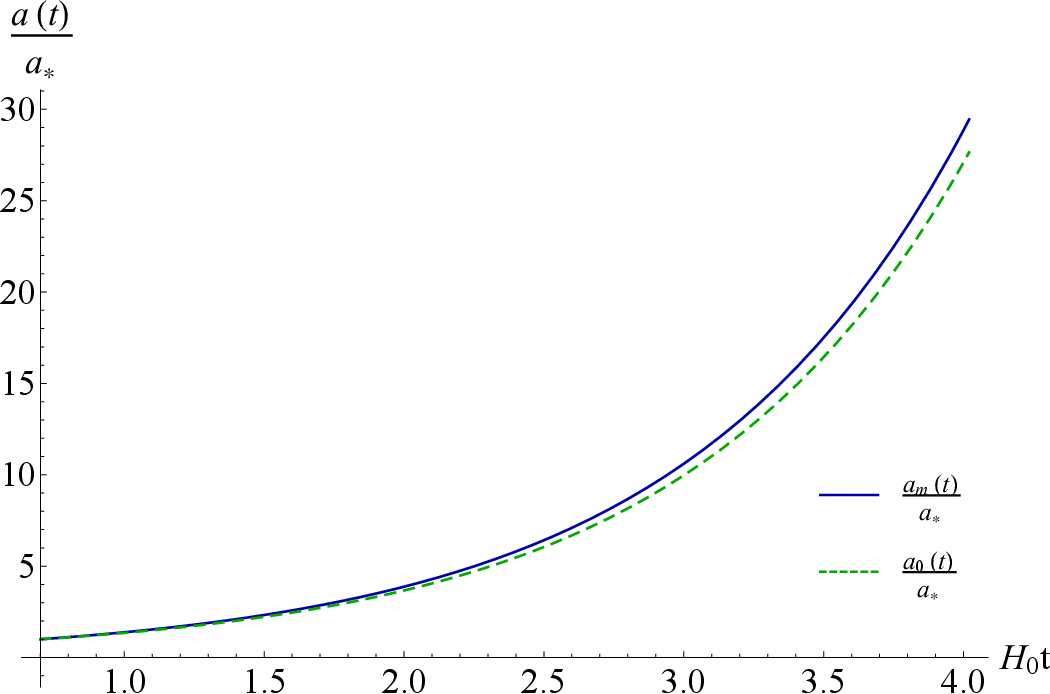}
\caption{Scale factor versus time plot after the fixed point regime\label{Fig9}.}
\end{center}
\end{figure}

\noindent We continue with the model proposed in section (\ref{S4.B}) which indeed is the most general form of the dark-fluid considering a stiff matter-like, radiation-like, and matter-like components. In this regime, $a>a_*$ and as a result $\xi>1$. Using the form of $k$ from eq.(\ref{4.B.50}) and substituting it back in eq.(\ref{5.69}), we can write down the modified Friedmann equation outside of the fixed point regime as
\begin{widetext}
\begin{equation}\label{5.A.72}
\begin{split}
\frac{H^2}{H_0^2}\simeq&1+\frac{2\nu}{\omega\Lambda_0}\left(\frac{8\pi g_*\varepsilon_*}{\lambda_*}\right)^{\frac{1}{2}}\left(\frac{\Omega_{\text{m},*}}{\left(\nicefrac{a}{a_*}\right)^3}+\frac{\Omega_{\text{rad},*}}{\left(\nicefrac{a}{a_*}\right)^4}+\frac{\Omega_{\text{s},*}}{\left(\nicefrac{a}{a_*}\right)^6}\right)^\frac{1}{2}-\frac{\nu G_0}{\Lambda_0}\left(\frac{8\pi g_*\varepsilon_*}{\lambda_*}\right)\left(\frac{\Omega_{\text{m},*}}{\left(\nicefrac{a}{a_*}\right)^3}+\frac{\Omega_{\text{rad},*}}{\left(\nicefrac{a}{a_*}\right)^4}+\frac{\Omega_{\text{s},*}}{\left(\nicefrac{a}{a_*}\right)^6}\right)~.
\end{split}
\end{equation} 
\end{widetext}
As in this regime $k$ decreases which implies $a$ increases rapidly we can get rid of any higher order contributions which helps us to write down the above equation as
\begin{equation}\label{5.A.73}
\frac{H^2}{H_0^2}\simeq1+\frac{2\nu}{\omega\Lambda_0}\left(\frac{8\pi g_*\varepsilon_*}{\lambda_*}\right)^{\frac{1}{2}}\left(\frac{\Omega_{\text{m},*}}{\left(\nicefrac{a}{a_*}\right)^3}+\frac{\Omega_{\text{rad},*}}{\left(\nicefrac{a}{a_*}\right)^4}\right)^\frac{1}{2}~.
\end{equation}
In the above equation, it can be easily seen that the second term in the right hand side is sub-leading contribution compared to the first term as $a>a_*$ in this regime. One can keep the stiff-matter contribution which is equivalent to a $\xi^{-3}$ order contribution and in that case one also needs to keep the $-\frac{\nu G_0}{\Lambda_0}(\frac{8\pi g_*\varepsilon_*}{\lambda_*})\frac{\Omega_{\text{m},*}}{\xi^3}$ term from eq.(\ref{5.A.72}). Taking squareroot of the both sides of eq.(\ref{5.A.73}), we can approximately recast the equation as
\begin{equation}\label{5.A.74}
H\simeq H_0+\frac{\nu H_0}{\omega \Lambda_0}\left(\frac{8\pi g_*\varepsilon_*}{\lambda_*}\right)^{\frac{1}{2}}\left(\frac{\Omega_{\text{m},*}}{\left(\nicefrac{a}{a_*}\right)^3}+\frac{\Omega_{\text{rad},*}}{\left(\nicefrac{a}{a_*}\right)^4}\right)^\frac{1}{2}~.
\end{equation}
We can further simplify the above relation by dropping the radiation contribution as well leading to a simpler form for Eq.(\ref{5.A.74}) given by
\begin{equation}\label{5.A.74.B}
H\simeq H_0+\frac{\nu H_0}{\omega \Lambda_0}\left(\frac{8\pi \Omega_{\text{m},*} g_*\varepsilon_*}{\lambda_*}\right)^{\frac{1}{2}}\frac{1}{\left(\nicefrac{a}{a_*}\right)^\frac{3}{2}}~.
\end{equation}
The reason behind keeping the final term in eq.(\ref{5.A.74.B}) lies in the fact that in this perturbative regime $\frac{a}{a_*}\geq1$, which implies the only terms with lowest powers of $\frac{a}{a_*}$ in the denominator will contribute to the right hand side of eq.(\ref{5.A.72}). It is evident from eq.(\ref{5.A.74.B}) that in the perturbative regime only the matter part dominates. As has been discussed after eq.(\ref{4.B.51}), there is no inflation in the early time  for a renormalization group improved cosmology. As $\Lambda_0$ describes the value of the cosmological constant in the current time, therefore eq.(\ref{5.71}) denotes a dark energy dominated current-time accelerated expansion for the universe. Hence, an inflation is absent in the early time for a renormalization group improved cosmological model. As a result the quantum gravity corrections presented in eq.(\ref{5.A.74.B}), although small, will not wash away completely. One can now simplify eq.(\ref{5.A.74.B}) as
\begin{equation}\label{5.A.74.C}
\begin{split}
\frac{\frac{d\left(\nicefrac{a}{a_*}\right)}{dt}}{\left(\nicefrac{a}{a_*}\right)}&=H_0\left(1+\frac{\nu}{\omega\Lambda_0}\left(\frac{8\pi \Omega_{\text{m},*} g_*\varepsilon_*}{\lambda_*}\right)^{\frac{1}{2}}\frac{1}{\left(\nicefrac{a}{a_*}\right)^\frac{3}{2}}\right)\\
\implies H_0 dt&=\frac{\left(\nicefrac{a}{a_*}\right)^{\frac{1}{2}}d\left(\nicefrac{a}{a_*}\right)}{\left(\nicefrac{a}{a_*}\right)^\frac{3}{2}+\frac{\nu}{\omega\Lambda_0}\left(\frac{8\pi \Omega_{\text{m},*} g_*\varepsilon_*}{\lambda_*}\right)^{\frac{1}{2}}}~.
\end{split}
\end{equation} 
We now make a change of variables (following the early time analysis) as $\xi\equiv\frac{a}{a_*}$ which implies that at the transition point $t=t_*$, $\xi=1$ as $a(t_*)=a_*$. Integrating both sides of the above equation from the transition point at $t=t_*$ to some arbitrary $t$ such that $t>t_*$ , we get
\begin{equation}\label{5.A.75}
H_0(t-t_*)\simeq\int_1^\xi\frac{d\xi~\xi^\frac{1}{2}}{\xi^\frac{3}{2}+\frac{\nu}{\omega\Lambda_0}\left(\frac{8\pi g_*\varepsilon_*\Omega_{\text{m},*}}{\lambda_*}\right)^\frac{1}{2}}~.
\end{equation}
Defining $\sigma_{\text{m}}\equiv\frac{\nu}{\omega\Lambda_0}\left(\frac{8\pi g_*\varepsilon_*\Omega_{\text{m},*}}{\lambda_*}\right)^\frac{1}{2}$, we can simplify the above equation as
\begin{equation}\label{5.A.76}
\begin{split}
H_0(t-t_*)&=\frac{2}{3}\ln\left[\xi^\frac{3}{2}+\sigma_{\text{m}}\right]-\frac{2}{3}\ln\left[1+\sigma_{\text{m}}\right]\\
\implies \frac{a_{\text{m}}(t)}{a_*}&=\left[(1+\sigma_{\text{m}})e^{\frac{3H_0 (t-t_*)}{2}}-\sigma_{\text{m}}\right]^\frac{2}{3}.
\end{split}
\end{equation}
As $\sigma_{\text{m}}$ is a quantum gravity correction, it is very small and therefore for $t\gg t_*$, we can simplify the above equation as $\frac{a_\text{m}(t)}{a_*}\simeq\left(1+\frac{2\sigma}{3}\right)e^{H_0(t-t_*)}$.
One can use the values of $g_*$ and $\lambda_*$ as $g_*\simeq 0.27$, and $\lambda_*\simeq0.36$ \cite{BonanoReuter2}. Now as discussed in \cite{BonanoReuter2},  $\nu$ and $\omega$ are both positive numbers. At the fixed point $k\simeq m_{pl}\propto\frac{1}{\sqrt{G_0}}$ (in natural units), which helps us to approximately estimate $\varepsilon_*$ from eq.(\ref{2.20}) as
\begin{equation}\label{5.A.77}
\varepsilon_*\simeq \frac{\nu}{4\pi\omega G_0^2}.
\end{equation}
We can use the dark-matter parameter $\Omega_{\text{m},*}$ to be 0.1 as have we used in sections (\ref{S4.B},\ref{S4.D}) to plot the scale factors and energy densities against time.
Instead of using proper values for the parameter, one is quite safe with the consideration $\sigma_{\text{m}}<1$. For our consideration we have used $\sigma_{\text{m}}=0.1$. We now plot $\frac{a_\text{m}(t)}{a_*}$ against $H_0t$ for the current case against the case with only dark-energy contribution in Fig.(\ref{Fig9}).
\begin{figure}
\begin{center}
\includegraphics[scale=0.48]{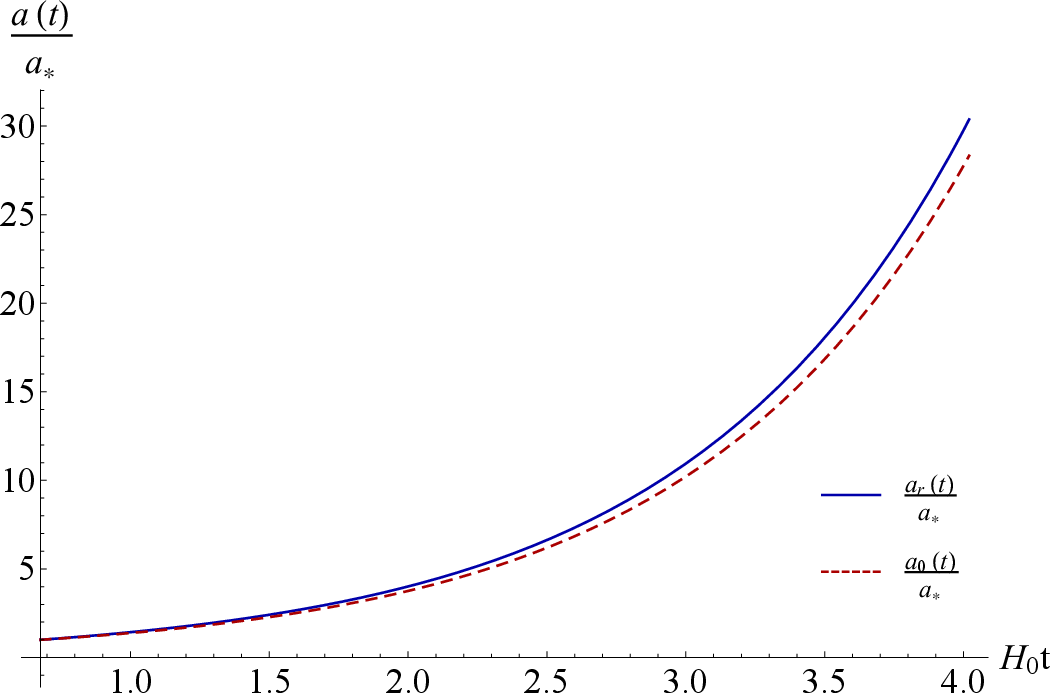}
\caption{Scale factor versus time plot after the fixed point regime when no dark-matter era is present\label{Fig10}.}
\end{center}
\end{figure}
In Fig.(\ref{Fig9}), $a_0(t)$ represents the solution in eq.(\ref{5.71}) and $a(t)$ represents the solution in eq.(\ref{5.A.76}). We observe that the dark matter dominated era leads to certain changes in the late-time behaviour of the universe where we have set $H_0t_*\simeq0.70$. It is important to understand that if $\sigma_{\text{m}}$ is made sufficiently small, then even for large times the dark matter dominated era will not change the overall dynamics of the evolution of the universe. If there is no dark matter dominated era present, then eq.(\ref{5.A.74}) can be recast in the form
\begin{equation}\label{5.A.78}
H\simeq H_0+\frac{\nu H_0}{\omega \Lambda_0}\left(\frac{8\pi g_*\varepsilon_*}{\lambda_*}\right)^{\frac{1}{2}}\left(\frac{\Omega_{\text{rad},*}}{\left(\nicefrac{a}{a_*}\right)^4}\right)^\frac{1}{2}~.
\end{equation}
Doing an integration of both sides of the above equation, we obtain 
\begin{equation}\label{5.A.79}
\begin{split}
H_0(t-t_*)\simeq&\int_1^\xi \frac{d\xi~\xi}{\xi^2+\sigma_\text{r}}\\
=&\frac{1}{2}\left(\ln[\xi^2+\sigma_\text{r}]-\ln[1+\sigma_\text{r}]\right)
\end{split}
\end{equation}
where $\sigma_{\text{r}}$ is defined as
\begin{equation}\label{5.A.80}
\sigma_{\text{r}}\equiv\frac{\nu}{\omega\Lambda_0}\left(\frac{8\pi g_*\varepsilon_*\Omega_{\text{rad},*}}{\lambda_*}\right)^\frac{1}{2}.
\end{equation}
Solving the integral in eq.(\ref{5.A.79}), one can obtain the time dependent solution of $a_{\text{r}}(t)$ as
\begin{equation}\label{5.A.81}
a_{\text{r}}(t)\simeq a_*\left((1+\sigma_{\text{r}})e^{2H_0(t-t_*)}-\sigma_\text{r}\right)^\frac{1}{2}.
\end{equation}
We plot $\frac{a_{\text{r}}(t)}{a_*}$ against $\frac{a_0(t)}{a_*}$ versus $H_0t$ in Fig.(\ref{Fig10}) where we have set $\sigma_\text{r}=0.15$ and $H_0t_*\simeq0.68$. Again we observe that a radiation dominated era results in a change in the long time behaviour of the accelerated expansion of the universe which is same as the case as has been observed for the matter dominated era. In the next subsection, we consider the behaviour of the scale factor after the fixed point regime when considering the case of section(\ref{S4.D}) with an universe filled with stiff-matter+anti-radiation+dark-matter fluid. One important thing to remember about the scale factor plots is that before the fixed point regime one should not plot upto the fixed point which stands same for the current case. For the sake of continuity of the result, we have plotted upto and from the fixed point for section(\ref{S4}) and section(\ref{S5}) respectively.
\section{Possibility of cyclic solution of the universe in the late-time regime for a RG improved cosmology}\label{S5B}
\noindent In this section, we shall investigate whether it is possible to obtain a cyclic or bouncing kind of solution of the universe for a particular combination of the dark fluid as has been observed for standard FLRW cosmological model in \cite{Chavanis0}. Now in our analysis, we mainly focussed on renormalization group improved cosmology in the early as well as the late time. It is evident from our analyses that obtaining a bouncing-like solution with components with positive coefficients only is not possible. Again RG-improved cosmology with very high energy density before the Planck-time or transition time $t_*$ does not allow a bouncing structure. We shall therefore concentrate only for the cases where the energy density is relatively lower and this helps us to stay well inside the perturbative regime.
As we are considering the perturbative (as well as the late time regime), we can simply recast eq.(\ref{5.69}) as
\begin{equation}\label{6.1}
H^2\simeq \frac{\Lambda_0}{3}+\frac{2\nu k^2}{3\omega}
\end{equation}
where we have, for now, got rid of the $\mathcal{O}(k^4)$ contribution. As, we are in the low energy regime, we can use the identification of the energy density from eq.(\ref{2.20}) and substitute it back in the above equation, which allows us to recast eq.(\ref{6.1}) as
\begin{equation}\label{6.2}
H^2=\frac{\Lambda_0}{3}+\frac{8\pi G_0\varepsilon}{3}
\end{equation}
which is the standard Friedmann equation without contribution from the renormalization group improvement. In order to observe correction to the late time behaviour of the scale factor one also needs to consider the final term in  eq.(\ref{5.69}) which has been dropped in eq.(s)(\ref{6.1},\ref{6.2}).
\subsection{Stiff matter, matter and anti-dark energy}\label{6.A}
\begin{figure}[ht!]
\begin{center}
\includegraphics[scale=0.245]{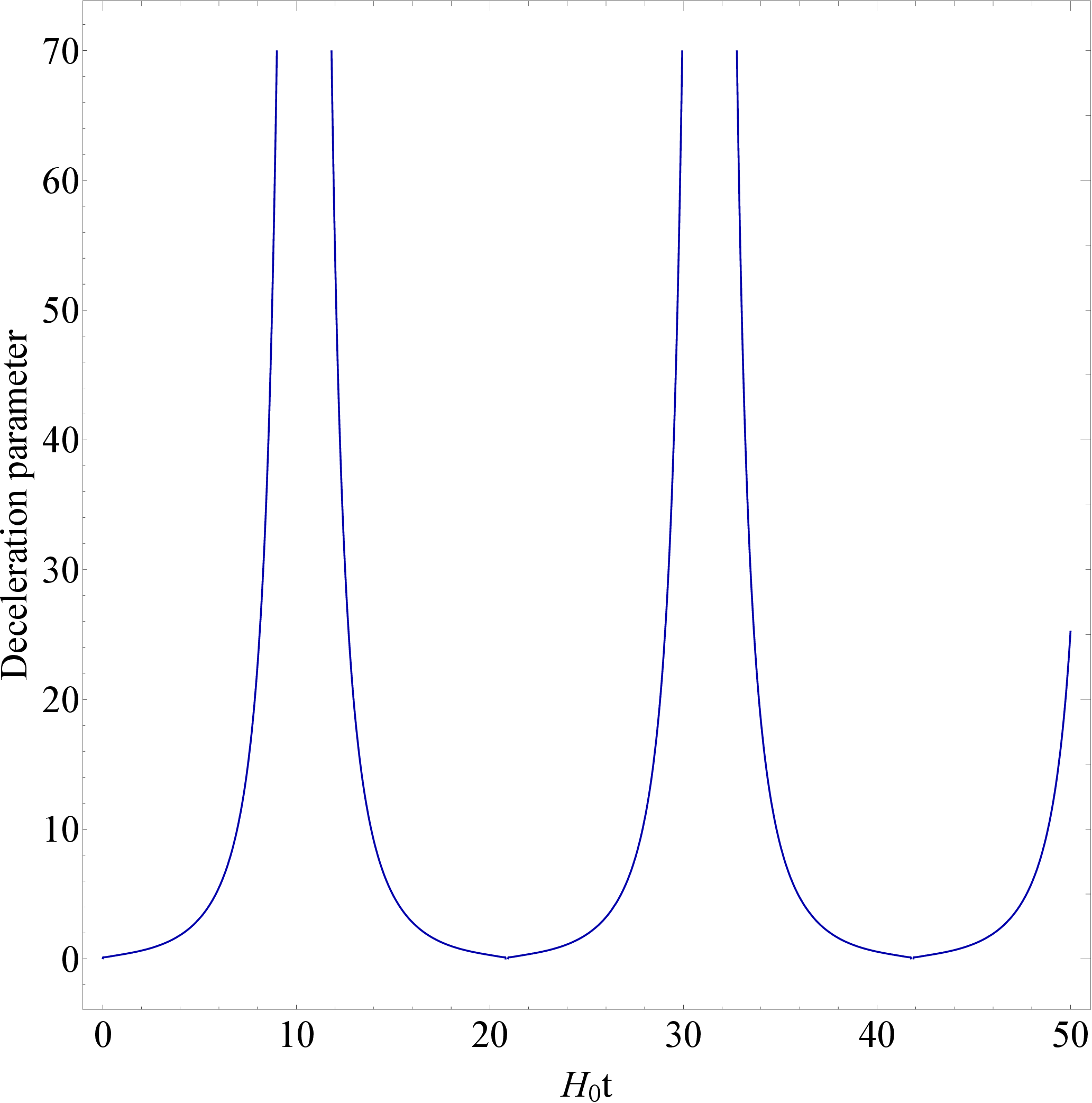}
\caption{Plot of $a_d(t)$ vs $H_0 t$ for a RG improved cosmology with stiff-matter, dark-matter, and anti-dark energy components.\label{Deceleration}}
\end{center}
\end{figure}

\noindent We start by considering that the dark fluid is a combination of a negative dark energy component and dark matter component.
The Friedmann equation in eq.(\ref{6.2}) then takes the form
\begin{equation}\label{6.3}
H^2=-\frac{|\Lambda_0|}{3}+\frac{8\pi G_0\varepsilon_0}{3}\left(\frac{\Omega_{\text{m} ,0}}{\left(\frac{a}{a_0}\right)^3}+\frac{\Omega_{\text{s},0}}{\left(\frac{a}{a_0}\right)^6}\right)~.
\end{equation}
Now the above equation is considered primarily after the Planck time. As we are primarily interested in the late-time behaviour, we have set the reference point at the present time of our universe which is $t=t_0$. The energy density $\varepsilon=0$ for 
\begin{equation}\label{6.4}
\begin{split}
\frac{a_F}{a_0}=\left(\frac{\Omega_{\text{m},0}+\sqrt{\Omega_{\text{m},0}^2+4\Omega_{\text{s},0}|\Omega_{\Lambda,0}|}}{2\Omega_{\Lambda,0}}\right)^\frac{1}{3}
\end{split}
\end{equation}
where $|\Omega_{\Lambda,0}|=\frac{|\Lambda_0|}{8\pi G_0\varepsilon_0}$.
A careful analysis reveals the analytical form of the dimensionless scale factor to be
\begin{equation}\label{6.5}
\begin{split}
\frac{a}{a_0}=&\left(\frac{\Omega_{\text{m},0}}{|\Omega_{\Lambda,0}|}\sin^2\left(\frac{3}{2}\sqrt{|\Omega_{\Lambda,0}|}H_0(t-t_*)\right)\right.\\&+\left.\sqrt{\frac{\Omega_{\text{s},0}}{|\Omega_{\Lambda,0}|}}\sin\left(3 \sqrt{|\Omega_{\Lambda_0}|}H_0 (t-t_*)\right)\right)^{\frac{1}{3}}+\frac{a_*}{a_0}
\end{split}
\end{equation}
where proper choice for the constants have been made to obtain the above solution and in the above equation $t_*$ can always be neglected for a late time scenario and $\frac{a_*}{a_0}$ is added by hand as one can set the integration constants depending on the initial condition. For a late time analysis the above result is identical to the analysis conducted in \cite{Chavanis0}. Substituting the form of the scale factor in the energy density expression, one obtains the time at which the scale factor reaches its maximum value $a=a_F$ to be
\begin{equation}\label{6.6}
t_F=t_*+\frac{1}{3\sqrt{|\Omega_{\Lambda,0}|}H_0}\left(\pi-\tan^{-1}\left(\frac{2\sqrt{\Omega_{\text{s},0}|\Omega_{\Lambda,0}|}}{\Omega_{\text{m},0}}\right)\right)~.
\end{equation}
At this point the energy density of the universe vanishes. Again at $t=3t_F-2t_*$, the energy density of the universe vanishes indicating a bouncing kind of solution which is very similar to the case observed in \cite{Chavanis0}. In the $0\leq t\leq t_*$, regime the stiff matter component will dominate the expansion of the universe. It is more prudent to start the analysis in the fixed point regime but as we are more interested  in the late time behaviour we have ignored the time evolution of the scale factor in the $0\leq t\leq t_*$ regime. One can obtain the deceleration parameter for this particular case which is given by the expression
\begin{equation}\label{6.7}
a_d(t)\equiv-\frac{\frac{\ddot{a}(t)}{a_0}}{H^2(t)}~.
\end{equation}
For this current analysis, the analytical form of the deceleration parameter is obtained as
\begin{widetext}
\begin{equation}\label{6.8}
a_d(t)=\frac{\frac{\mathcal{A}(t)}{16}\left(\frac{\Omega_{\text{m},0}}{|\Omega_{\Lambda,0}|}\left(1-\cos\left(3H_\Lambda(t-t_*)\right)\right)+2\sqrt{\frac{\Omega_{\text{s},0}}{|\Omega_{\Lambda,0}|}}\sin\left(3H_\Lambda(t-t_*)\right)\right)}{\left[\frac{\Omega_{\text{m},0}}{|\Omega_{\Lambda,0}|}\sin^2\left(\frac{3}{2}H_\Lambda(t-t_*)\right)+\sqrt{\frac{\Omega_{\text{s},0}}{|\Omega_{\Lambda,0}|}}\sin\left(3H_\Lambda(t-t_*)\right)\right]^\frac{2}{3}\left[\sqrt{\frac{\Omega_{\text{s},0}}{|\Omega_{\Lambda,0}|}}\cos(3H_\Lambda (t-t_*))+\frac{\Omega_{\text{m},0}}{|\Omega_{\Lambda,0}|}\sin(3H_\Lambda (t-t_*))\right]^2}
\end{equation}
where $H_\Lambda\equiv\sqrt{|\Omega_{\Lambda,0}|}H_0$ and the analytical form of $\mathcal{A}(t)$ is given by
\begin{equation}\label{6.9}
\begin{split}
\mathcal{A}(t)&=5\left(\frac{\Omega^2_{\text{m},0}}{|\Omega_{\Lambda,0}|^2}+\frac{4\Omega_{\text{s},0}}{|\Omega_{\Lambda,0}|}\right)-\frac{6\Omega^2_{\text{m},0}}{|\Omega_{\Lambda,0}|^2}\cos(3H_\Lambda(t-t_*))+\left(\frac{\Omega^2_{\text{m},0}}{|\Omega_{\Lambda,0}|^2}-\frac{4\Omega_{\text{s},0}}{|\Omega_{\Lambda,0}|}\right)\cos(6H_\Lambda (t-t_*))\\
&-\frac{4\Omega_{\text{m},0}}{|\Omega_{\Lambda,0}|}\sqrt{\frac{\Omega_{\text{s},0}}{|\Omega_{\Lambda,0}|}}\left(\sin(6H_\Lambda(t-t_*))-3\sin(3H_\Lambda(t-t_*))\right)~.
\end{split}
\end{equation}
\end{widetext}
We shall now plot the deceleration parameter with respect to the dimensionless time $H_0 t$ in Fig.(\ref{Deceleration}). The coefficient values are set as $\Omega_{\text{m},0}=0.1$, $|\Omega_{\Lambda,0}|=0.01$, $\Omega_{\text{s},0}=0.0001$, and $H_0t_*=0.0001$.
 From the plot of the deceleration parameter in Fig.(\ref{Deceleration}), it is evident that the universe indeed is bouncing.
Similar bouncing solutions can be observed for the combination of anti-stiff matter, matter and anti-dark energy. In the next subsection, we investigate the case of the dark fluid consist of stiff matter and anti-dark matter.
\subsection{Dark matter and anti-dark energy}
\noindent Now for a universe with a dark-matter and anti-dark energy like component, we shall write down the Friedmann equation upto $\mathcal{O}(k^4)$ which in terms of the energy density of the universe can be expressed as
\begin{equation}\label{6.2.1}
\begin{split}
H^2&=-\frac{|\Lambda_0|}{3}+\frac{8\pi G_0\varepsilon}{3}-\frac{16\pi^2G_0^3\omega^2\varepsilon^2}{\nu}\\
&=\frac{8\pi G_0\varepsilon_0}{3}\left[\frac{\Omega_{\text{m},0}}{\left(\frac{a}{a_0}\right)^3}-|\Omega_{\Lambda,0}|\right]-\frac{16\pi^2G_0^3\omega^2\varepsilon_0^2}{\nu}\frac{\Omega_{\text{m},0}^2}{\left(\frac{a}{a_0}\right)^6}~.
\end{split}
\end{equation}
This corrected form is very interesting in a sense that if we identify $|\Omega_{\text{s},0}^{\nu,\omega}|=\frac{6\pi \omega^2G_0^2\Omega_{\text{m},0}^2\varepsilon_0}{\nu}$. Then it is possible to express the above equation in a simpler form given as
\begin{equation}\label{6.2.1}
\frac{H^2}{H_0^2}=-\frac{|\Omega_{s,0}^{\nu,\omega}|}{\left(\frac{a}{a_0}\right)^6}+\frac{\Omega_{\text{m},0}}{\left(\frac{a}{a_0}\right)^3}-|\Omega_{\Lambda,0}|~.
\end{equation}
The above equation is very interesting in a sense that the $\mathcal{O}(k^4)$ term overall contributes in an anti-stiff matter like component in the Friedmann equation. One can solve the above equation exactly and the result will be similar to that of the results obtained in \cite{Chavanis0} for the case anti-stiff matter, dark matter, and anti-dark energy case which can be written simply as
\begin{equation}\label{6.2.2}
\frac{a}{a_0}\simeq\left[\frac{\Omega_{\text{m},0}}{2|\Omega_{\Lambda,0}|}-\frac{\sqrt{\Delta^{\nu,\omega}}}{2|\Omega_{\Lambda,0}|}\cos\left(3\sqrt{|\Omega_{\Lambda,0}|}H_0 (t-t_*)\right)\right]^\frac{1}{3}
\end{equation}
where $\frac{a_*}{a_0}$ contribution has been ignored as it is very small. In the above equation $\Delta^{\nu,\omega}$ is defined as
\begin{equation}\label{6.2.3}
\Delta^{\nu,\omega}=\Omega_{\text{m},0}^2-4|\Omega_{\Lambda,0}||\Omega_{\text{s},0}^{\nu,\omega}|~.
\end{equation}
It is always possible to take out the $|\Omega_{\text{s},0}^{\nu,\omega}|$ component from eq.(\ref{6.2.2}) and write a perturbative solution but for our current case, we restrain ourselves from doing that as it unnecessarily complicates the analytical structure of the scale factor. 
It is easy to observe that the above scale factor in eq.(\ref{6.2.2}) indicates towards a bouncing behaviour of the universe. The scale factor becomes minimum at $t=t_*$ and the scale factor becomes maximum at
\begin{equation}\label{6.2.4}
t_F=t_*+\frac{\pi}{3\sqrt{|\Omega_{\Lambda,0}|H_0}}~.
\end{equation}
Again at $t=2t_F-t_*$ the scale factor becomes minimum and this continues\footnote{Keep in my mind that the value of the scale factor at $t=t_*$ has been neglected as it is indeed very small when compared to the value of scale factor at late times.}. Similar behaviour can be observed for the case of the universe filled with Anti-stiff matter, dark matter and anti-dark energy. Only thing that would be interesting is that we need not define any fixed time $t_*$ as the universe never approaches infinite energy density as like all of the other models. As a result the UV-fixed point analysis of the RG improved universe becocmes unnecessary. It is although important to note if we have not considered any $\mathcal{O}(k^4)$ correction in the Friedmann equation, then for $t=t_*$ the scale factor in eq.(\ref{6.2.2}) vanishes indicating that the universe collapses to a singularity. Hence, the RG improvement prevents the universe to collapse to a singularity rendering a non-singular bouncing solution of the universe.
\section{Polytropic equation of state and inflation}\label{S6}
\noindent As we have see5n in the earlier analysis, an inflationary phase of the universe cannot exist before the Planck time as in order for the universe to go through inflation, one needs to have a phase where $a(t)$ grows rapidly (exponentially) with time. Inflation can only happen in the later part of the evolution which is after the Planck time and before the reheating era. Now previously we were getting accelerated expansion in the perturbative regime but only for the existence of the cosmological constant which implies the current accelerated expansion. In our current analysis, we set $\Lambda_0=0$ in eq.(\ref{5.69}) to rule out the possibility of an accelerated expansion of the universe. Hence, neglecting the $\mathcal{O}(k^4)$ term, it is quite plausible to write the Friedmann equation in the perturbative regime as
\begin{equation}\label{Inf.1}
H^2\simeq \frac{2\nu k^2}{3\omega }~.
\end{equation}
Before moving forward to the identification of $k$, we start by considering an equation of state as \cite{Chavanis0,Chavanis1,Chavanis2,Chavanis3}
\begin{equation}\label{Inf.2}
P=\alpha\varepsilon-(\alpha+1)\varepsilon \left(\frac{\varepsilon}{\rho_P}\right)^{\frac{1}{n}}
\end{equation}
where $0\leq\alpha\leq1$, $n>0$, and $\rho_P=\frac{1}{G_0^2}$ denoting the Planck density. The above equation of state is a sum of two equations of state. The first one is a linear equation of state given by the relation $P=\alpha\varepsilon$ and the second one is a polytropic equation of state given by $P\propto \varepsilon^\gamma$ with $\gamma=1+\frac{1}{n}$. As discussed in \cite{Chavanis0}, the above equation of state in eq.(\ref{Inf.2}) for $\alpha=1$, shows a transition between inflation and a stiff matter era. Similarly $\alpha=\frac{1}{3}$ denotes the transition between a inflation and a radiation-dominated era.

\noindent Substituting the above equation of state in the continuity equation, one can obtain the form of the energy density as
\begin{equation}\label{Inf.3}
\varepsilon=\frac{\rho_{P}}{\left(1+\left(\nicefrac{a}{a^{(c)}}\right)^\frac{3(1+\alpha)}{n}\right)^n}
\end{equation}
where $a^{(c)}$ denotes an integration constant. It is important to remember that at the Planck time $\varepsilon\simeq \rho_P$ indicating that at this point $a\ll a^{(c)}$. This implies that not only $a_*\neq a^{(c)}$ but also $a_*\ll a^{(c)}$. As can be seen from eq.(\ref{Inf.3}), for $a<a^{(c)}$, the energy density is a constant ($\varepsilon\sim \rho_P$) which indicates an inflationary phase whereas for $a>a^{(c)}$, with $\alpha=1,\frac{1}{3},0$, and $n=1$, one expects a transition to stiff-matter, radiation and matter dominated eras respectively from the inflationary phase. The fact that the equation of state enables a transition between the inflation era to the different eras based on the value of $a$ with respect to the transition point $a^{(c)}$, is a motivation well enough for the choice of eq.(\ref{Inf.2}) as a physically consistent and valid equation of state.

\begin{figure}
\begin{center}
\includegraphics[scale=0.48]{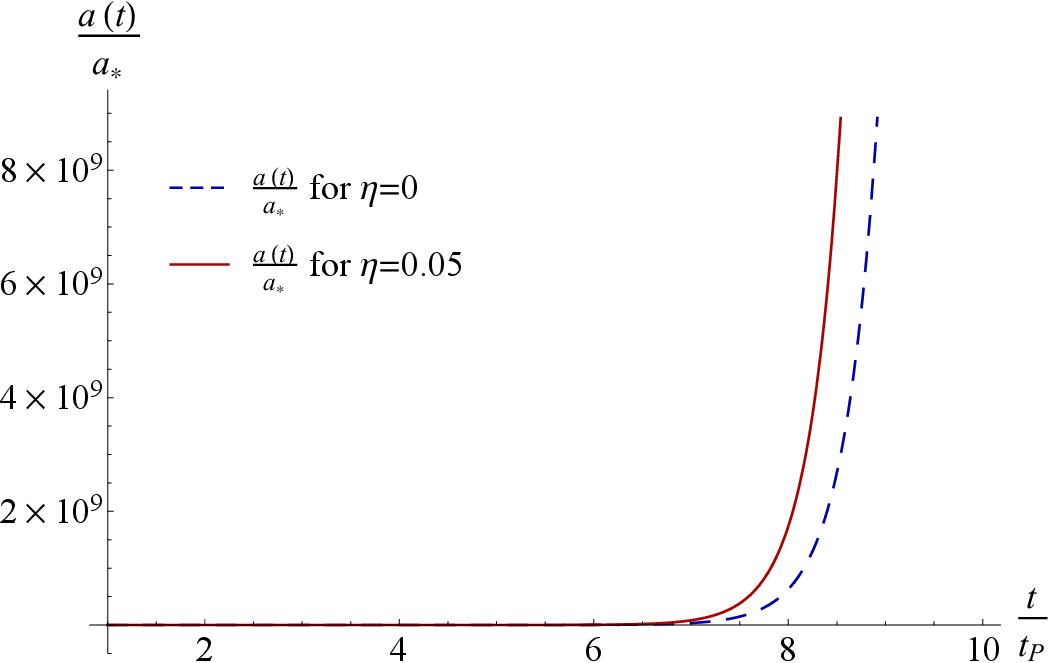}
\caption{Plot of the dimensionless scale factor against the dimensionless time of the universe in the context of inflation when no quantum gravity correction is present ($\eta=0$) versus the case when quantum gravity correction is present ($\eta=0.05$)\label{Fig_Inflation}.}
\end{center}
\end{figure}

\noindent Comparing the above equation with eq.(\ref{2.20}), we obtain
\begin{equation}\label{Inf.4}
k\simeq\left(\frac{4\pi\omega G_0 \rho_P}{\nu}\right)^\frac{1}{2}\frac{1}{\left(1+\left(\nicefrac{a}{a^{(c)}}\right)^\frac{3(1+\alpha)}{n}\right)^\frac{n}{2}}
\end{equation}
Taking squareroot of the both sides of eq.(\ref{Inf.1}) and substituting $k$ from the above equation, we obtain
\begin{equation}\label{Inf.5}
H\simeq\left(\frac{8\pi G_0\rho_p}{3}\right)^\frac{1}{2}\frac{1}{\left(1+\left(\nicefrac{a}{a^{(c)}}\right)^\frac{3(1+\alpha)}{n}\right)^\frac{n}{2}}~.
\end{equation}
The above result is quite interesting as  above result  is identical to the one obtained in \cite{Chavanis0}. This serves as a consistency check for the renormalization group improved cosmological model with the standard cosmological model discussed in \cite{Chavanis0}. Using the identification $t_{\text{Pl}}=\frac{1}{\sqrt{G_0\rho_p}}$, the general solution of eq.(\ref{Inf.5}) for $\alpha=n=1$ reads
\begin{widetext}
\begin{equation}\label{Inf.6}
\begin{split}
&\sqrt{\frac{8\pi }{3}}\frac{t-t_{\text{Pl}}}{t_{\text{Pl}}}=\frac{2}{3(1+\alpha)}\left[\sqrt{1+\xi_c^{3(1+\alpha)}}+\ln\left[\frac{\xi_c^{\frac{3(1+\alpha)}{2}}}{1+\sqrt{1+\xi_c^{3(1+\alpha)}}}\right]\right]\Biggr\rvert_{\tilde{\xi}_c}^{\xi_c}
\end{split}
\end{equation}
where $\xi_c$ and $\tilde{\xi}_c$ are defined as
\begin{equation}\label{Inf.7}
\xi_c=\frac{a}{a^{(c)}}~,~~\tilde{\xi}_c=\frac{a_*}{a^{(c)}}~.
\end{equation}
As $a_*\ll a^{(c)}$ or $\tilde{\xi}_c\ll1$ one can recast eq.(\ref{Inf.6}) as 
\begin{equation}\label{Inf.8}
\begin{split}
&\frac{3(1+\alpha)}{2}\sqrt{\frac{8\pi }{3}}\frac{t-t_*}{t_{\text{Pl}}}\simeq\sqrt{1+\xi_c^{3(1+\alpha)}}-1+\ln\left[\left(\frac{a(t)}{a_*}\right)^{\frac{3(1+\alpha)}{2}}\right]-\ln\left[\frac{1+\sqrt{1+\xi_c^{3(1+\alpha)}}}{2}\right]
\end{split}
\end{equation}
\end{widetext}
where we have substituted $\nicefrac{\xi_c}{\tilde{\xi}_c}=\nicefrac{a(t)}{a_*}$. It is important note that we can easily set $t_*=t_{\text{Pl}}$ in the above equation. Now for $a\ll a^{(c)}$, one can solve the above equation and obtain the time dependence of the scale factor as
\begin{equation}\label{Inf.9}
a\simeq a_* e^{\sqrt{\frac{8\pi\omega}{3}}\frac{t-t_{\text{Pl}}}{t_{\text{Pl}}}}~.
\end{equation}
The fundamental difference of this result with the one presented in \cite{Chavanis0} is that in our case the solution is not valid before the Planck time which forces us to implement a lower bound of integration for the time part to be $t_{\text{Pl}}$ and the scale factor part to be $a_*$. For $a\gg a^{(c)}$, we arrive at the solution
\begin{equation}\label{Inf.10}
\frac{a(t)}{a^{(c)}}\simeq\left(\frac{3(1+\alpha)}{2}\sqrt{\frac{8\pi }{3}}\frac{t-t_{\text{Pl}}}{t_{\text{Pl}}}\right)^\frac{2}{3(1+\alpha)}.
\end{equation}
In this regime, the energy density and scale factor are no more constant and have an algebraic time evolution which for $\alpha=1$ denotes the stiff matter era. 
%To investigate the effect of the flow parameter $\omega$, we plot the dimensionless scale factor $\frac{a}{a_*}$ against $\frac{t}{t_{\text{Pl}}}$ for different values of $\omega$ ($\omega={1,0.95}$) in Fig.(\ref{Inflation}).
%\begin{figure}
%\begin{center}
%\includegraphics[scale=0.45]{Inflation.eps}
%\caption{We plot $\frac{a(t)}{a_*}$ versus $\frac{t}{t_{\text{Pl}}}$ for $\omega=\{1,0.95\}$.\label{Inflation}}
%\end{center}
%\end{figure}
%From Fig.(\ref{Inflation}), it is evident that in the beyond fixed point regime the current model gives an inflationary behaviour. We also find out that for smaller values of the flow parameter $\omega$, the rate of inflation is delayed.
In order to truly investigate the effects of the flow parameters in the inflationary behaviour, we need to take $\mathcal{O}(k^4)$ term in eq.(\ref{5.69}). The Friedmann equation with $\Lambda_0=0$ gives
\begin{equation}\label{Inf.11}
H^2=\frac{2\nu k^2}{3\omega}-\frac{\nu G_0 k^4}{3}~.
\end{equation}
Substituting the value of $k$ from eq.(\ref{Inf.4}) in the above equation and taking a square root, we obtain the following relation
\begin{equation}\label{Inf.12}
H=\frac{\left(\frac{8\pi G_0\rho_p}{3}\right)^\frac{1}{2}}{\left(1+\xi_c^\frac{3(1+\alpha)}{n}\right)^\frac{n}{2}}\left(1-\frac{2\pi \rho_p\omega^2 G_0^2}{\nu\left(1+\xi_c^{\frac{3(1+\alpha)}{n}}\right)^n}\right)^\frac{1}{2}~.
\end{equation}
From eq.(s)(\ref{2.16},\ref{2.17}), we can see that both $\omega$ and $\nu$ are correction parameters in the late time expansion of $G(k)$ and $\Lambda(k)$ respectively. Again $G_0^2$ itself is a very small quantity and as a result we can treat $\frac{\pi\rho_p\omega^2G_0^2}{\nu}$ as a small quantity and define it by $\eta$. Hence, we can recast eq.(\ref{Inf.12}) (with the assumption that $\eta\ll 1$) as
\begin{equation}\label{Inf.13}
H\simeq\frac{\left(\frac{8\pi G_0\rho_p}{3}\right)^\frac{1}{2}}{\left(1+\xi_c^\frac{3(1+\alpha)}{n}\right)^\frac{n}{2}}\left(1-\frac{\eta}{\left(1+\xi_c^{\frac{3(1+\alpha)}{n}}\right)^n}\right)~.
\end{equation} 
We shall now look into the case for when $\alpha=1$ and $n=1$. We can further recast the above equation in the form given as
\begin{equation}\label{Inf.14}
\begin{split}
\int_{\tilde{\xi}_c}^{\xi_c} \frac{d\xi_c}{\xi_c}\left(\sqrt{1+\xi^6}+\frac{\eta}{\sqrt{1+\xi^6}}\right)&=\sqrt{\frac{8\pi}{3}}\int_{t_{\text{Pl}}}^t\frac{dt}{t_{\text{Pl}}}\\
&=\sqrt{\frac{8\pi}{3}}\frac{t-t_{\text{Pl}}}{t_{\text{Pl}}}~.
\end{split}
\end{equation}
Performing the integral in the left hand side, we obtain the following relation
\begin{widetext}
\begin{equation}\label{Inf.15}
\begin{split}
\sqrt{\frac{8\pi}{3}}\frac{t-t_{\text{Pl}}}{t_{\text{Pl}}}=&\frac{1}{3}\left(\sqrt{1+\xi_c^6}-\sqrt{1+\tilde{\xi}_c^6}\right)+(1-\eta)\ln\left[\nicefrac{\xi_c}{\tilde{\xi}_c}\right]-\frac{1-\eta}{3}\ln\left[\frac{\sqrt{1+\xi_c^6}+1}{\sqrt{1+\tilde{\xi}_c^6}+1}\right]~.
\end{split}
\end{equation}
\end{widetext}
Now, as per we have discussed earlier, $\tilde{\xi}_c\ll 1$. If we again approach the regime when $\xi_c\rightarrow\tilde{\xi}_c$ or $a\ll a^{(c)}$, we arrive at the simplified expression
\begin{equation}\label{Inf.16}
\begin{split}
\sqrt{\frac{8\pi}{3}}\frac{t-t_{\text{Pl}}}{t_{\text{Pl}}}&\simeq (1-\eta)\ln\left[\frac{a}{a_*}\right]\\
\implies a(t)&\simeq a_*\exp\left[\frac{8\pi}{3}(1+\eta)\left(\frac{t-t_{\text{Pl}}}{t_{\text{Pl}}}\right)\right]~.
\end{split}
\end{equation}
Some important observations are in order. We observe from the above equation that the exponential evolution of the scale factor is independent of the value of the cosmological constant at the present time (we have already set $\Lambda_0=0$) which indicates the absence of any dark energy like term.This rules out the possibility of this time evolution to be identified with the accelerated expansion of the universe in the very late time regime and rather reinforces an inflationary dynamics after the Planck time regime in a renormalization group improved cosmology. As we can see that $\eta=\frac{\pi \rho_P\omega^2G_0^2}{\nu}$ is a purely quantum gravitational correction and we find that the quantum gravitational correction amplifies the inflation as the time increases with respect to the Planck time. The important aspect in eq.(\ref{Inf.16}) is that with the increase in the value of the relative time parameter $\frac{t}{t_{Pl}}$, the effect of the $\eta$ correction will be significant. Instead of substituting the values of the individual parameters, we consider an overall small value for $\eta$ and plot it against time where we compare the general case in \cite{Chavanis0} along with the  quantum gravity modified inflationary scenario in Fig.(\ref{Fig_Inflation}).
From Fig.(\ref{Fig_Inflation}), it is straightforward to check that the quantum gravity correction induces a higher rate of inflation corresponding to the case when $\eta=0$ after the Planck time. The inflation continues up to the time $t=t^{(c)}$ which can be considered as the starting of the reheating era. The important thing to observe is that the value of $\eta$ is quite small for such a significant amount of change in the rate of inflation. As the inflation is enhanced, the classical value of the scale factor in the beginning of the reheating era will be approached faster for a renormalized group improved cosmological model. Now we shall investigate the general structure of the energy density required for having an inflationary model which can be thought of as a back calculation. We assume that the energy density has the form given by
\begin{equation}\label{Inf.N.1}
\varepsilon=\frac{\varepsilon_P}{\left(1+\kappa\left[\frac{a}{a^{(c)}}\right]^j\right)^l}
\end{equation}
where $\kappa,~i,~j\in\mathbb{R}$. 
Substituting the above equation in the continuity equation in eq.(\ref{2.13a}), we can obtain the equation of state as
\begin{equation}\label{Inf.N.2}
P=\frac{\varepsilon_P}{\left(1+\kappa\left[\frac{a}{a^{(c)}}\right]^j\right)^{l+1}}\left(\kappa\left[\frac{a}{a^{(c)}}\right]^j\left(\frac{jl}{3}-1\right)-1\right)~.
\end{equation}
From eq.(\ref{Inf.N.1}), one can obtain the expression of the dimensionless scale factor in terms of the energy density as
\begin{equation}\label{Inf.N.3}
\frac{a}{a^{(c)}}=\left(\left(\frac{\varepsilon}{\varepsilon_P}\right)^{-\frac{1}{l}}-1\right)^{\frac{1}{j}}~.
\end{equation}
Substituting, eq.(\ref{Inf.N.3}) back in eq.(\ref{Inf.N.2}), we get the generalized equation of state as
\begin{equation}\label{Inf.N.4}
P=\left(\frac{jl}{3}-1\right)\varepsilon-\frac{jl\varepsilon}{3}\left(\frac{\varepsilon}{\varepsilon_P}\right)^{\frac{1}{l}}~.
\end{equation}
As in \cite{Chavanis0} as well as in our analysis the inflationary behaviour is dominant in the early phase when $a\ll a^{c}$, then only $\varepsilon\sim \varepsilon_P$ which is a constant. This leads to $k\propto \sqrt{\varepsilon_P}$ and $H=\text{constant}$. As a result, one obtains an inflationary behaviour when $a\ll a^{(c)}$. One can also propose an energy density of the form $\varepsilon=\varepsilon_P\left(1\pm \kappa'\left(\frac{a}{a^{(c)}}\right)^{j'}\right)^{l'}$\footnote{For the mentioned form of the energy density the equation of state simply reads $P=-\left(1+\frac{j'l'}{3}\right)\varepsilon+\frac{j'l'\varepsilon}{3}\left(\frac{\varepsilon}{\varepsilon_P}\right)^{-\frac{1}{l'}}$. If $j'<0$ and $l'>0$ such that $j'l'<-3$ then we get non-trivial equation of states where the energy density is of the form $\varepsilon=\varepsilon_P\left(1\pm \frac{\kappa'}{\left(\frac{a}{a^{(c)}}\right)^{|j'|}}\right)^{l'}$. Now if one takes $j'>0$ and $l'<0$ such that $j'l'<-3$ then we get back the form similar to eq.(\ref{Inf.N.1}).} kind of structure with the only requirement being $\varepsilon\sim \text{constant}$ when $a\ll a^{(c)}$.  

\subsection{Inflation followed by a pre-Planckian stiff-matter era}
\begin{figure}
\begin{center}
\includegraphics[scale=0.48]{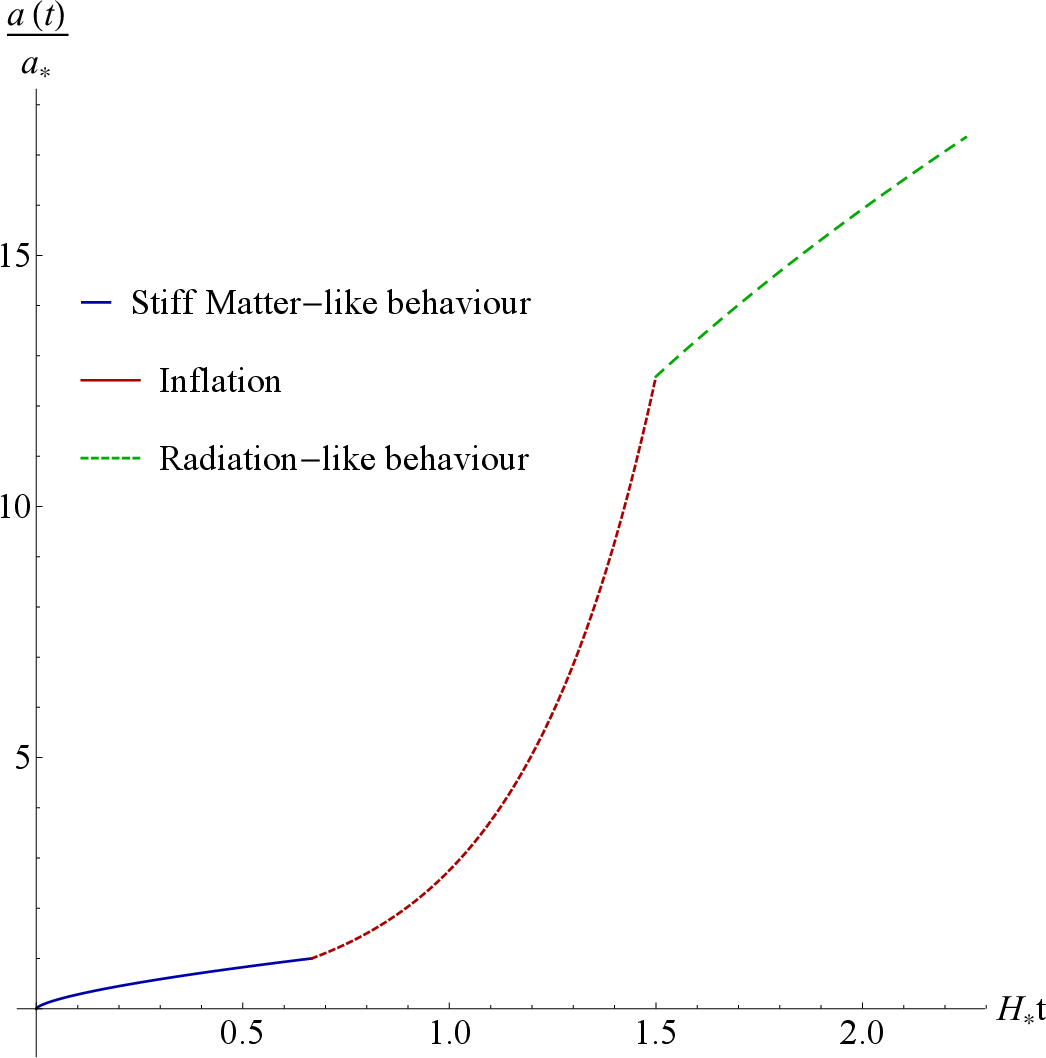}
\caption{A complete evolution of the universe is presented in the pre nucleosynthesis era. \label{Fig_Inflation_complete}}
\end{center}
\end{figure}

\noindent The form of the energy density in eq.(\ref{Inf.3}), mainly depends on the fact that there is a transition between an inflation era to a stiff-matter era after the Planck time (provided $\alpha=n=1$ in eq.(\ref{Inf.3})). We instead propose a more physical form of the energy density given as
\begin{equation}\label{Inf.N.1}
\begin{split}
\varepsilon=\frac{\rho_P}{\left(1+\left(\nicefrac{a}{a^{(c)}}\right)^{\frac{3(1+\alpha)}{n}}\right)^n}+\frac{\rho_P^S}{\left(\nicefrac{a}{a_*}\right)^6}~.
\end{split}
\end{equation}
The form of the energy density in the above equation comes from a pressure term (in terms of the scale factor of the universe) given by
\begin{equation}\label{Inf.N.2}
P=\frac{\rho_P^S}{\left[\frac{a}{a_*}\right]^6}+\frac{\alpha\rho_P}{\left[1+\left[\frac{a}{a^{(c)}}\right]^{\frac{3(1+\alpha)}{n}}\right]^n}-\frac{(1+\alpha)\rho_P}{\left[1+\left[\frac{a}{a^{(c)}}\right]^{\frac{3(1+\alpha)}{n}}\right]^{n+1}}.
\end{equation}
The primary difference of the energy density in eq.(\ref{Inf.16}) with the one in eq.(\ref{Inf.N.1}) is that in eq.(\ref{Inf.N.1}) there is an additional $\frac{\rho_P^S}{\left[\frac{a}{a_*}\right]^6}$ term which imitates a dominant stiff matter contribution before $t=t_*$.
The form of the energy density in eq.(\ref{Inf.N.1}) has very interesting effects and imitates three major evolution periods in the universe. 
\begin{enumerate}
\item For $a(t)<a_* ~(\ll a^{(c)})$, we observe a stiff-matter dominated early expansion of the universe. It is imprtant to note that this stiff matter dominate early ``power-law" expansion dominates only up to the Planck-time.
\item For $a_*\leq a(t)\leq a^{(c)}$, we obtain an inflationary phase and therefore, the equation of state in eq.(\ref{Inf.N.1}) gives rise to a transition from the stiff matter era to an inflationary phase which is in contrast to the one given by eq.(\ref{Inf.16}) where the stiff-matter dominated phase comes after the inflation.
\item When $a(t)>a^{(c)}$, we observe a transition to a radiation-like or dark matter-like behaviour depending on the value of $\alpha$ and $n$. 
\end{enumerate}
The coefficient of the stiff-matter component in eq.(\ref{Inf.N.1}) can be identified to $\rho_P^S\equiv\Omega_{\text{s},*}\varepsilon_*$ which can be set equals to $\rho_P$ for the sake of analytical convenience. As the post-inflationary reheating phase is considered to be mainly dominated by radiation, we consider $\alpha=\frac{1}{3}$ and $n=1$ in eq.(\ref{Inf.N.1}). As has been extensively studied in section (\ref{S4}), one should identify $k$ using the analytical form of $\varepsilon(k)$ inside the fixed point regime. The standard analytical procedure gives $a(t)\propto t^{\frac{2}{3}}$ inside the fixed point regime. Then in the $t_*\leq t\leq t^{(c)}$ regime ($t^{(c)}$ denotes the transition time between the inflationary period and the re-heating era), $a(t)$ gives the exponential growth obtained in eq.(\ref{Inf.16}). Finally, in the $t>t^{(c)}$ regime we obtain $a(t)\propto t^{\frac{1}{2}}$ which indicates a radiation-dominated ``power-law"expansion of the universe. We now plot the complete evolution of the dimensionless  scale factor of the universe against the dimensionless time before the nucleosynthesis era in Fig.(\ref{Fig_Inflation_complete}). We have used a  very small time interval $t^{(c)}-t_*$, for the inflationary period to highlight the nature of time-dependence of the scale factor in the three-segments. This above model is physically sound and explains that a primitive stiff-matter era before the fixed-time regime has a contribution towards the expansion of the universe.
\section{Conclusion}\label{SConclusion}
\noindent In this paper, we consider a cosmological model where the universe has gone through a stiff matter era. Instead of using the standard cosmological model, we make use of the renormalization group improved cosmology by considering the flow of the Newton's gravitational constant as well as the cosmological constant with the momentum scale of the universe $k$. The $k$ dependence of $G(k)$ and $\Lambda(k)$ are different when one considers the epoch the universe before the Planck time (fixed point scaling regime) and when one considers after the Planck time (perturbative regime). At first we investigate the existence of a stiff-matter fluid in the fixed point scaling regime along with preferred combinations of a dark matter, radiation, and anti-radiation era. It is important to note that the universe is considered to be filled with a dark fluid which behaves as different components during different times of the evolution period of the universe. We now introduce a new approach where we identify the momentum scale of the universe $k$ with the scale factor via comparing the $k$ dependent form of the energy density in the $k>m_{Pl}$ regime and scale factor dependent form of the energy density. We have then substituted the $a(t)$ dependent form of $k$ in the Friedmann equation and obtained a time-dependent solution of the scale factor. Plotting the dimensionless scale factor with change in the dimensionless time, we find out that for $a(t)\ll a_*$ the stiff matter era dominates and results in a general accelerated expansion of the universe upto the Planck time. We observe in the section (\ref{S4.D}), that a stiff matter era along with just anti-radiation will lead to several unexpected behaviour which we argued will be the same for a stiff matter together with anti-dark matter component as well. In order get a good behaviour of the evolution of the universe before the Planck time, one needs to consider a stiff matter+anti-radiation+dark matter combination or a stiff matter+radiation+anti-dark matter combination. One important aspect to note that unlike \cite{Chavanis0}, the current model will break down for a anti-stiff matter component consideration as the starting energy density then will be zero which is in contradiction to the renormalization group improved approach at very high energy scales. Although it is possible to still use the perturbative Friedmann equations as has been discussed in section (\ref{S5B}). We also argued that before the Planck time, it is impossible to have an inflation era in the renormalization group framework. Next, we have investigated the beyond Planck time regime for the dark fluid to consist of a stiff matter+radiation components, and stiff matter+radiation+dark matter components. We observe that in this regime where $a>a_*$, the dark matter part has the subleading order contribution where the leading order contribution to the accelerated expansion of the universe comes from the current value of the cosmological constant. In the absence of a dark matter component the radiation part has the next sub-leading contribution towards the evolution of the scale factor. In section (\ref{S5B}), we have discussed few bouncing solutions of the universe in the existence of an anti-dark energy component in the dark fluid. Finally, in section(\ref{S6}), we have used the polytropic equation of state introduced in \cite{Chavanis0,Chavanis1,Chavanis2,Chavanis3} and observed that in the beyond Planck time regime (perturbative regime), one can get inflation and the inflation dynamics is similar to the one observed in \cite{Chavanis0}. The important observation is that the inflationary behaviour comes out from the $k^2$ order correction term of the Friedmann equation in eq.(\ref{5.69}). This is an important result in our work. We have also computed quantum gravity corrections to the inflationary behaviour of the scale factor which indicates that inflation is enhanced by such effects. Next, we have proposed the general class of equations of state that result in an inflationary model in the renormalization group improved cosmology as well as the general cosmological model discussed in \cite{Chavanis0}. Finally, we have proposed a new energy-density which results in a very early stiff-matter era (in the fixed point scaling regime) and then it introduces an inflation era just after the Planck time which exists upto the beginning of the reheating phase at $t=t^{(c)}$. After $t^{(c)}$  a radiation-dominated era begins which extends upto the nucleosynthesis epoch. We believe that the energy-density introduced in eq.(\ref{Inf.N.1}) gives the most realistic universal evolution model for a renormalization group improved cosmological scenario.
\appendix
\section{Cosmology with negative energy components}\label{A1}
\noindent For academic completeness of our analysis, we now consider some negative-energy components in the evolution of cosmology before and after fixed time era. It is although quite known among the cosmologists that no traces of negative energy components are observed in the current time. In principle positive energy components lead to cosmological expansion of the universe whereas negative energy components primarily result in a contraction of the cosmos. It is quite important to know that the existence of negative-energy components after the fixed point regime are quite niche as it is quite unlikely to contribute to meaningful expansion of the universe. Although before the fixed point regime the existence of anti-particle states like anti-radiation can be a  possibility which has been diluted as other positive energy states overtook its effect with evolution of time. A detailed analysis of the effect of negative energy densitites in Friedmann cosmology can be found in \cite{Patla}. From the perspective of string theory a fundamental problem is the appearence of innumerous number of vacua.  It turns out that the existence of negative energy density components will result in a substantial deduction of the existing number vacua in the theory \cite{AntiEnergy}.
\subsection{Anti radiation and stiff matter era before the Planck time}\label{S4.C}
\begin{figure}
\begin{center}
\includegraphics[scale=0.48]{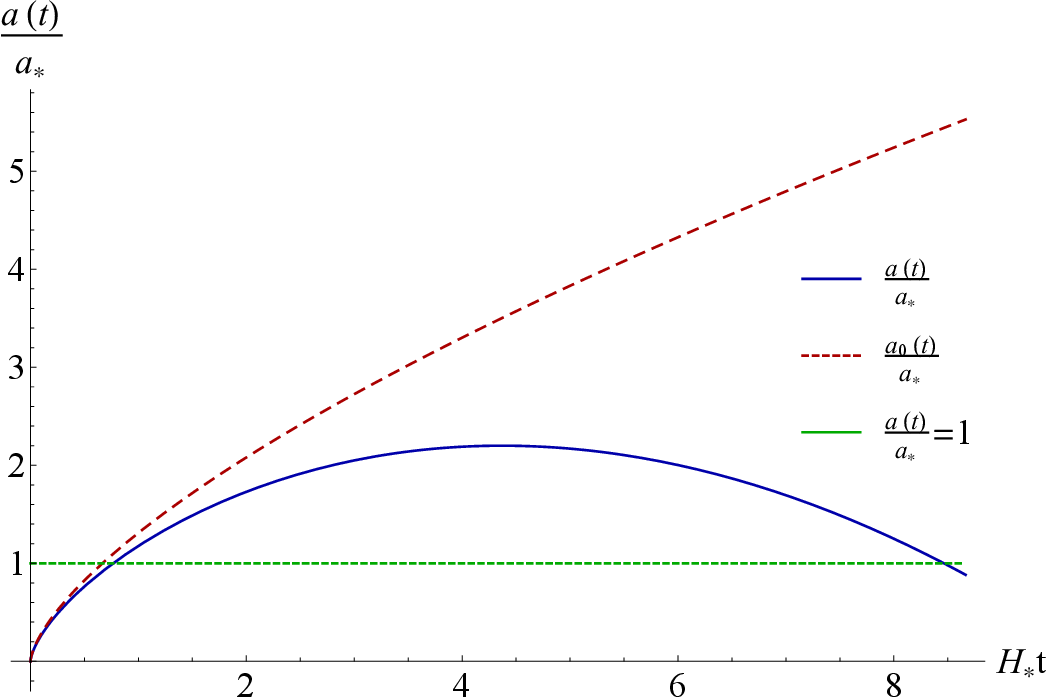}
\caption{Plot of $\frac{a(t)}{a_*}$ versus $H_*t$ for Stiff-matter+anti-radiation era against the stiff-matter only\label{Fig5}.}
\end{center}
\end{figure}

\begin{figure}
\begin{center}
\includegraphics[scale=0.48]{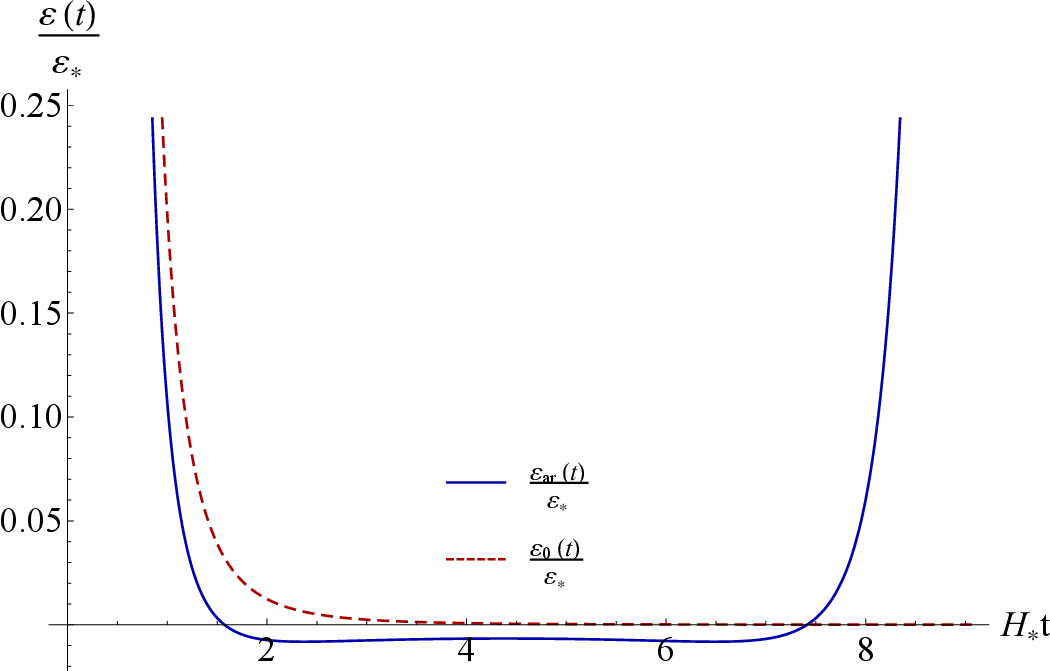}
\caption{Energy density versus time plot for stiff-matter+anti-radiation era against the only stiff matter era.\label{Fig6}}
\end{center}
\end{figure}

\noindent Here, we discuss that in the very early universe there was stiff-matter era followed by an anti-radiation fluid which indicates that the modified energy density takes the form 
\begin{equation}\label{4.C.58}
\begin{split}
\frac{\varepsilon}{\varepsilon_*}=\frac{\Omega_{\text{s},*}}{\left(\nicefrac{a}{a_*}\right)^6}-\frac{|\Omega_{\text{arad},*}|}{\left(\nicefrac{a}{a_*}\right)^4}
\end{split}
\end{equation}
with $\Omega_{\text{arad},*}$ denoting the energy-density corresponding to the anti-radiation part divided by the energy-density at the fixed time limit. The primarly problem of working with negative energy densities is that it may lead to the viloation of local energy conditions. As has been discussed in deltails in several extensions of quantum field theories \cite{QFTViolation,QFTViolation2,QFTViolation3,
QFTViolation4,QFTViolation5,QFTViolation6,
QFTViolation7}, relativistic energy conditions only need to satisfy globally. This allows one for small local violations as has also been discussed in \cite{Patla}. It is therefore possible that before the Planck time there had been a component with negative energy density which had contributed to the expansion of the universe for a very short period of time. It is still important from an analytical perspective to implement beforehand the fact that $\Omega_{\text{s},*}>\Omega_{\text{arad},*}$ as the first term in the right hand side of the above equation shall dominate the second term for very small values of the scale factor. If we claim that $\varepsilon=0$, then eq.(\ref{4.C.58}) gives that
\begin{equation}\label{4.C.59}
\frac{a}{a_*}=\sqrt{\frac{\Omega_{\text{s},*}}{|\Omega_{\text{arad},*}|}}~.
\end{equation}
As long as $\Omega_{\text{s},*}>\Omega_{\text{arad},*}$ condition holds, the above equation suggests that for $\varepsilon$ to be zero, $a>a_*$ which is an impossible condition as $a<a_*$ inside the fixed point regime. This contradiction shows that inside the fixed point regime, $\varepsilon$ can never be zero let alone be negative provided there is stiff-matter domination in this regime. As a result the minimum value of $a$ can still be zero. Again using eq.(\ref{2.15}) and comparing it with eq.(\ref{4.C.58}), we arrive at the identification of the momentum scale $k$ with the scale factor $a$ as
\begin{equation}\label{4.C.60}
k=\left(\frac{8\pi g_*\varepsilon_*}{\lambda_*}\right)^\frac{1}{4}\left(\frac{\Omega_{\text{s},*}}{\left(\nicefrac{a}{a_*}\right)^6}-\frac{|\Omega_{\text{arad},*}|}{\left(\nicefrac{a}{a_*}\right)^4}\right)^\frac{1}{4}.
\end{equation}
Using the above equation, we can write down the Friedmann equation as
\begin{equation}\label{4.C.61}
\frac{H^2}{H_{*}^2}=\left(\frac{\Omega_{\text{s},*}}{\left(\nicefrac{a}{a_*}\right)^6}-\frac{|\Omega_{\text{arad},*}|}{\left(\nicefrac{a}{a_*}\right)^4}\right)^\frac{1}{2}.
\end{equation}
Again defining the same variable $\xi$ as $\frac{a}{a_*}$, we get from the Friedmann equation after an integration of both sides as
\begin{equation}\label{4.C.62}
\begin{split}
H_*t&= \frac{2\xi^{\frac{3}{2}}\left(\Omega_{\text{s},*}-\xi^2|\Omega_{\text{arad},*}|\right)^\frac{3}{4}}{3\Omega_s}\tensor[_{2}]{F}{_{1}}\left[1,\frac{3}{2},\frac{7}{4},\frac{|\Omega_{\text{arad,*}}|\xi^2}{\Omega_{\text{s},*}}\right]\\
&\simeq \frac{2\xi^{\frac{3}{2}}}{3\Omega_{\text{s},*}^{\frac{1}{4}}}\left(1+\frac{3|\Omega_{\text{arad},*}|\xi^2}{28\Omega_{\text{s},*}}\right)~.
\end{split}
\end{equation}
A perturbative solution gives $a$ as a function of $t$ given by
\begin{equation}\label{4.C.63}
\frac{a(t)}{a_*}\simeq \left(\frac{3\Omega_{\text{s},*}^{\frac{1}{4}}H_*t}{2}\right)^\frac{2}{3}-\frac{|\Omega_{\text{arad},*}|}{14\Omega_{\text{s},*}}\left(\frac{3\Omega_{\text{s},*}^{\frac{1}{4}}H_*t}{2}\right)^2~.
\end{equation}
Unlike the ever-increasing nature of the scale factor with time for the previous cases, it is evident from the above equation that $a(t)$ will have a maxima after which the value of $a(t)$ starts to decrease. Defining a new quantity $\tau\equiv H_*t$ (the dimensionless time parameter), we can find the transition point by using the condition $\frac{d a(\tau)}{d\tau}=0$ to be
\begin{equation}\label{4.C.64}
\tau_c=\frac{2\sqrt{\Omega_{s,*}}}{3|\Omega_{\text{arad},*}|^\frac{3}{4}}\left(\frac{14}{3}\right)^\frac{3}{4}.
\end{equation} 
Beyond $\tau_c$, anti-radiation starts to dominate more than the stiff-matter part. It is though important to note that at $\tau_c$, $\frac{a}{a_*}>1$ which indicates that at the maxima we are beyond the validity of the current model. Plotting the scale factor for the above case against the stiff-matter only case, we observe the exact same behaviour leading to an unexpected nature of evolution of the scale factor. We now plot the dimensionless scale factor against the dimensionless time in Fig.(\ref{Fig5}).
We observe that for anti-radiation with a stiff-matter era the scale factor decreases after hitting a maxima at $\tau_c$ but again for a later time $a(t)$ becomes smaller than $a_*$. In our case, the used values of the parameters are $\Omega_{\text{s,*}}=0.7$ and $|\Omega_{\text{arad},*}|=0.3$ and we observe that the current model is valid for $H_*t\lesssim0.77$ and $H_*t\gtrsim 8.46$. This leads to certain discrepancies in the model. Plotting the dimensionless energy density against the dimensionless time in Fig.(\ref{Fig6}), we observe that the energy density becomes negative after a certain time and then again at a later time it becomes positive which implies that a stiff-matter along with a anti-radiation era cannot exist.
From Fig.(\ref{Fig6}) it can also be inferred that stiff-matter+anti-dark matter era cannot exist as well in the early universe. Although to implement such a solution one can set a cut-off value of the scale factor which is $a_f=a_*$. For the next subsection, we consider a stiff-matter+anti-radiation+dark-matter era.
\subsection{Stiff matter, anti radiation, and dark matter}\label{S4.D}
\begin{figure}
\begin{center}
\includegraphics[scale=0.48]{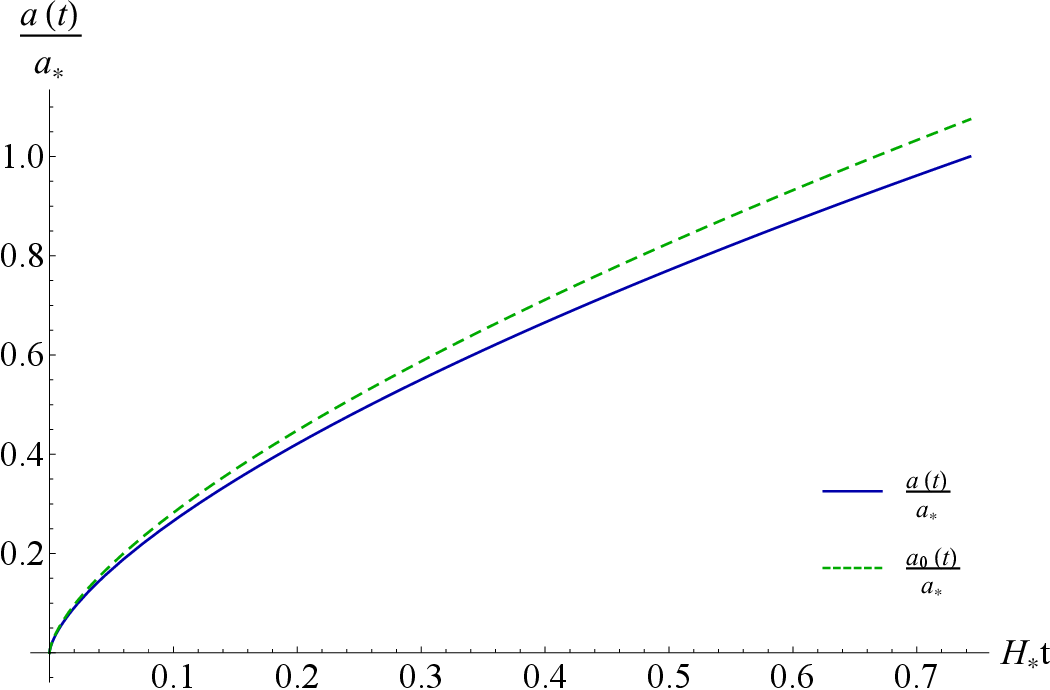}
\caption{We plot the scale factor for stiff-matter+anti-radiation+dark-matter dominated universe against the stiff-matter dominated universe when $t$ is varying.\label{Fig7}}
\end{center}
\end{figure}

\begin{figure}
\begin{center}
\includegraphics[scale=0.48]{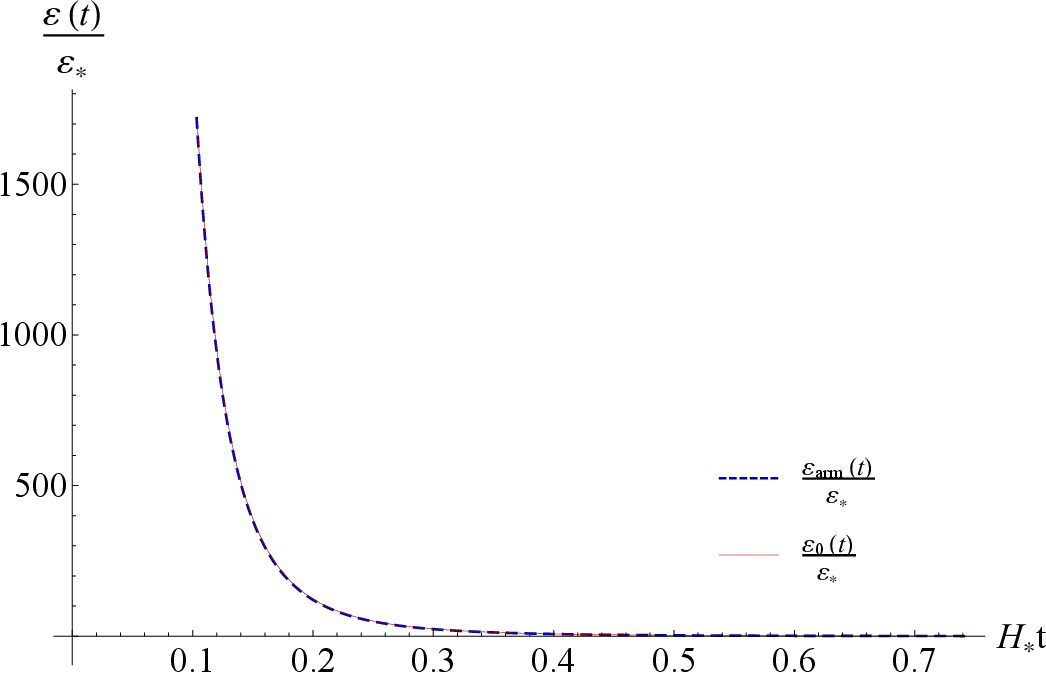}
\caption{Plot of energy density for the case of stiff-matter+anti-radiation+dark-matter against stiff-matter dominated universe while $t$ is varying\label{Fig8}.}
\end{center}
\end{figure}

\noindent For the stiff matter, anti radiation, and dark matter fluid filled early universe we can write down the energy density as
\begin{equation}\label{4.D.65}
\frac{\varepsilon}{\varepsilon_*}=\frac{\Omega_{\text{s},*}}{\left(\nicefrac{a}{a_*}\right)^6}-\frac{|\Omega_{\text{arad},*}|}{\left(\nicefrac{a}{a_*}\right)^4}+\frac{\Omega_{\text{m},*}}{\left(\nicefrac{a}{a_*}\right)^3}~.
\end{equation}
We can express the momentum scale in terms of the scale factor as
\begin{equation}\label{4.D.66}
k=\left(\frac{8\pi g_*\varepsilon_*}{\lambda_*}\right)^\frac{1}{4}\left(\frac{\Omega_{\text{s},*}}{\left(\nicefrac{a}{a_*}\right)^6}-\frac{|\Omega_{\text{arad},*}|}{\left(\nicefrac{a}{a_*}\right)^4}+\frac{\Omega_{\text{m},*}}{\left(\nicefrac{a}{a_*}\right)^3}\right)^\frac{1}{4}~.
\end{equation}
Following the similar steps, we get the equation among $\xi$ and $t$ as
\begin{equation}\label{4.D.67}
\begin{split}
H_* t=\frac{2\xi^{\frac{3}{2}}}{3\Omega_{\text{s},*}^{\frac{1}{4}}}\left(1+\frac{3|\Omega_{\text{arad},*}|\xi^2}{28\Omega_{\text{s},*}}-\frac{\Omega_{\text{m},*}\xi^{3}}{12\Omega_{\text{s},*}}\right)~.
\end{split}
\end{equation}
Solving the above equation perturbatively we arrive at the solution of $a(t)$ as a function of $t$ as
\begin{equation}\label{4.D.68}
\begin{split}
\frac{a(t)}{a_*}\simeq&\left(\frac{3\Omega_{\text{s},*}^{\frac{1}{4}}H_*t}{2}\right)^\frac{2}{3}-\frac{|\Omega_{\text{arad},*}|}{14\Omega_{\text{s},*}}\left(\frac{3\Omega_{\text{s},*}^{\frac{1}{4}}H_*t}{2}\right)^2\\&+\frac{\Omega_{\text{m},*}}{18\Omega_{\text{s},*}}\left(\frac{3\Omega_{\text{s},*}^{\frac{1}{4}}H_*t}{2}\right)^\frac{8}{3}~.
\end{split}
\end{equation}
In Fig.(\ref{Fig7}), we plot the scale factor in eq.(\ref{4.D.68}) for the current case against the scale factor for the universe with only stiff-matter domination inside the fixed point regime.
We find out from Fig.(\ref{Fig7}), that the scale factor has a slower rate of increase than the stiff matter only case but even at a very later time it never decreases so that $\frac{a(t)}{a_*}$ hits unity. The parameter values used for plotting eq.(\ref{4.D.68}) are $\Omega_{\text{s},*}=0.7$, $\Omega_{\text{rad},*}=0.2$, and $\Omega_{\text{m},*}=0.1$. Substituting the solution in eq.(\ref{4.D.68}) back in eq.(\ref{4.D.65}), we obtain the energy density for the stiff-matter+anti-radiation+dark-matter case\footnote{It is important to note that although we obtain a perturbative solution for $a(t)$ we do not inverse the result while plotting $\varepsilon$ as it leads to very minor contributions which do not contribute in much differences in the plots.}. We now plot the energy density for the current case which we define as $\varepsilon_{\text{arm}}(t)$ against $\varepsilon_0(t)$ while $t$ is varying in Fig.(\ref{Fig8}).
From the energy density plot in Fig.(\ref{Fig8}), we observe that the energy density in this current choice has a perfectly suitable behaviour unlike the previous case which makes it a better candidate for an early-universe model with flow than the preceding case. Now stiff-matter+radiation+anti-dark matter also shows similar behaviour to the current analysis and as a result we are not repeating the identical analysis. On the other hand stiff-matter+anti-radiation+anti-dark matter case also has the same phantom behaviour to the stiff-matter+anti-radiation model resulting in an unacceptable early universe model for renormalization group improved cosmology. One important aspect that can be observed from this section that one does not have an inflation era because of the fact that in the renormalization group improved Friedmann equation in eq.(\ref{4.36}), it is not possible to obtain a inflation era as inside the fixed point regime $H$ can not be made to be proportional to a constant which leads to an exponential inflation era inside or very near the fixed point regime. In the next section, we shall investigate the evolution equations for the time after the fixed point regime.
\subsection{Dark matter, anti radiation, and stiff matter era beyond the Planck time}\label{S5.B}
\noindent In section (\ref{S4.D}) we have investigated the early universe with an stiff matter+anti-radiation+dark matter era. We found out that the early time dynamics although is dominated mainly by the stiff matter era, the dynamics tends to deviate as an effect of the anti radiation era and the dark-matter era. Now, in this subsection we shall investigate the beyond fixed time implication for the same model (perturbative regime). With the identification of $k$ from eq.(\ref{4.D.66}) one can write down the  modified Friedmann equation from eq.(\ref{5.69}) for the perturbative regime as (we get similar expression from eq.(\ref{5.A.72}) where $\Omega_{\text{rad,*}}$ is replaced by $-\abs{\Omega_{\text{arad,*}}}$)
\begin{widetext}
\begin{equation}\label{S.B.82}
	\begin{split}
  \frac{H^2}{H_0^2}\simeq&1+\left(\frac{2\nu}{\omega\Lambda_0}\right)\left(\frac{8\pi g_{*}\varepsilon_{*}}{\lambda_{*}}\right)^\frac{1}{2}\left[\frac{\Omega_{\text{m},*}}{(a/a_*)^3}+\frac{\Omega_{\text{s},*}}{(a/a_*)^6}-\frac{\abs{\Omega_{\text{arad,*}}}}{(a/a_*)^4}\right]^\frac{1}{2}- \frac{\nu G_0}{\Lambda_0}\left(\frac{8\pi g_{*}\varepsilon_{*}}{\lambda_{*}}\right)\left[\frac{\Omega_{\text{m},*}}{(a/a_*)^3}+\frac{\Omega_{\text{s},*}}{(a/a_*)^6}-\frac{\abs{\Omega_{\text{arad,*}}}}{(a/a_*)^4}\right]
  \end{split}
\end{equation}
As we are working in the perturbative regime of the universe, where $k$ is low and  $a(t)$ is also very large than the fixed point scale factor $a_*$, it is safe to take the following approximation
\begin{equation}\label{S.B.83}
	\begin{split}
	\left(\frac{H}{H_0}\right)&\simeq1+\left(\frac{\nu}{\omega\Lambda_0}\right)\left(\frac{8\pi g_{*}\varepsilon_{*}}{\lambda_{*}}\right)^\frac{1}{2}\Omega_{\text{m},*}^\frac{1}{2}\left(\frac{a_*}{a}\right)^\frac{3}{2}\left(1
+\frac{\Omega_{\text{s},*}/\Omega_{\text{m},*}}{2(a/a_*)^3}-\frac{\left|\Omega_{\text{arad},*}\right|/\Omega_{\text{m},*}}{2(a/a_*)}\right)\\&-\frac{\nu G_0}{2\Lambda_0}\left(\frac{8\pi g_{*}\varepsilon_{*}}{\lambda_{*}}\right)\Omega_{\text{m},*}\left(\frac{a_*}{a}\right)^3\left(1+\frac{\Omega_{\text{s},*}/\Omega_{\text{m},*}}{(a/a_*)^3}-\frac{\left|\Omega_{\text{arad},*}\right|/\Omega_{\text{m},*}}{(a/a_*)}\right)+\mathbcal{O}\left(\left(\frac{a_*}{a}\right)^{13/2}\right)~.
\end{split}
\end{equation}
For simplification we shall take terms upto $(\frac{a_*}{a})^\frac{5}{2}$ and will drop higher order terms, implying 
\begin{equation}\label{S.B.84}
	\left(\frac{H}{H_0}\right)\simeq 1+ \left(\frac{\nu}{\omega\Lambda_0}\right)\left(\frac{8\pi g_{*}\varepsilon_{*}}{\lambda_{*}}\right)^\frac{1}{2}\Omega_{\text{m},*}^\frac{1}{2}(\frac{a_*}{a})^\frac{3}{2}\left(1
	-\frac{\left|\Omega_{\text{arad},*}\right|/\Omega_{\text{m},*}}{2(a/a_*)}\right)~.
\end{equation}
%$\bigr(\bigr)\Bigr(\Bigr)\biggr(\biggr)\Biggr(\Biggr)$
Performing integration of the both sides of the above equation, we get
\begin{equation}\label{S.B.85}
	H_0(t-t_*)\simeq\int_{1}^{\xi}\frac{d\xi}{\xi}\left(1+\frac{\nu}{\omega \Lambda_0}\left(\frac{8\pi \text{g}_*\varepsilon_*\Omega_{\text{m},*}}{\lambda_{*}}\right)^{1/2}\xi^{-3/2}-\frac{\nu\left|\Omega_{\text{arad},*}\right|}{2\omega \Lambda_0}\left(\frac{8\pi \text{g}_*\varepsilon_*}{\lambda_* \Omega_{\text{m}*}}\right)^{1/2}\xi^{-5/2}\right)^{-1}.
\end{equation}
Now this integration cannot be done exactly, but as the upper limit of the integration is greater than unity, therefore the second and the third term of the integrand is less dominating than the leading order term. Hence, we can expand the integrand followed by doing the integration to get the following equation (again keeping terms upto $\mathcal{O}(\xi^{-5/2})$ in the integrand)
\begin{equation}\label{SB.86}
H_0(t-t_*)\simeq\int_{1}^{\xi}\frac{d\xi}{\xi}\left(1-\frac{2\nu}{\omega \Lambda_0}\left(\frac{2\pi \text{g}_*\varepsilon_*\Omega_{\text{m},*}}{\lambda_{*}}\right)^{1/2}\xi^{-3/2}+\frac{\nu\left|\Omega_{\text{arad},*}\right|}{\omega \Lambda_0}\left(\frac{2\pi \text{g}_*\varepsilon_*}{\lambda_* \Omega_{\text{m}*}}\right)^{1/2}\xi^{-5/2}\right)~.
\end{equation}
\end{widetext}
Defining two new constants $\varsigma_{\text{m}}\equiv \frac{2\nu}{ \Lambda_0\omega}\sqrt{\frac{2\pi g_*\varepsilon_*\Omega_{\text{m}*}}{\lambda_{*}}}$ and $\varsigma_{\text{arm}}=\frac{\nu\left|\Omega_{\text{arad},*}\right|}{\Lambda_0\omega}\sqrt{\frac{2\pi g_*\varepsilon_*}{\lambda_{*}\Omega_{\text{m}*}}}$ we can recast eq.(\ref{SB.86}) after executing the integral as
%\begin{widetext}
\begin{equation}\label{S.B.87}
	H_0(t-t_*)\simeq ln(\xi)-\frac{2\varsigma_{\text{m}}}{3}(1-\xi^{-3/2})+\frac{2\varsigma_{\text{arm}}}{5}(1-\xi^{-5/2})~.
\end{equation}
For very late time the above equation further simplifies to  
\begin{equation}\label{S.B.88}
\begin{split}
	a(t)&\simeq a_*e^{H_0(t-t_*)+\frac{2\varsigma_{\text{m}}}{3}-\frac{2\varsigma_{\text{arm}}}{5}}
	\end{split}
\end{equation}
where we can express the above equation as $a(t)\simeq a_*e^{H_0(t-\tau)}$ when $\tau\equiv t_*-\frac{2}{3\text{H}_0}\left(\varsigma_{\text{m}}-\frac{3\varsigma_{\text{arm}}}{5}\right)$. 
Hence, it is clear to understand that the very late time dynamics completely agrees with the observed accelerated expansion of our Universe. As the plot of the scale factor have the exponential behaviour, we have not plotted $\frac{a(t)}{a_*}$ against $H_0t_*$ for the current case.

\section*{Acknowledgement}
\noindent G. G. extends his heartfelt gratitude to CSIR, Govt. of India for funding this research. 

\end{document}